\documentstyle[prl,aps,multicol,epsfig]{revtex}

\begin{document}
\large
\title{Spectral implementation of some quantum algorithms by one- and two-dimensional nuclear magnetic resonance}
\author{Ranabir Das$^\dagger$ and   
Anil Kumar $^{\dagger \ddagger}$\\
        $^{\dagger}$ {\small \it Department of Physics, Indian Institute of Science, Bangalore, India}\\
        $^{\ddagger}$ {\small \it Sophisticated Instruments Facility, Indian Institute of Science, Bangalore, India}\\}
\maketitle
\vspace{0.5cm}
\begin{abstract}
Quantum information processing has been effectively demonstrated on a small number of qubits by 
nuclear magnetic resonance. An important subroutine in any computing is the readout of the output. 
``Spectral implementation'' originally suggested by Z.L. Madi, R. Bruschweiler and R.R. Ernst,
 [J. Chem. Phys. 109, 10603 (1999)],
 provides an elegant method of readout with the use of an extra `observer' qubit.
At the end of computation, detection of the observer qubit provides the output 
via the multiplet structure of its spectrum. In ``spectral implementation" by two-dimensional experiment 
the observer qubit retains the memory of input state during computation, thereby 
providing correlated information on input and output, in the same spectrum. ``Spectral implementation" of Grover's search algorithm, 
 approximate quantum counting, a modified version of Berstein-Vazirani problem, and Hogg's algorithm  
is demonstrated here in three and four-qubit systems.
\end{abstract}

\section{Introduction}
 In 1982 Feynmann pointed out that it would be more efficient to simulate the behavior of  a quantum system
 using a quantum, rather than a classical device \cite{rf}. The idea of a purpose-built quantum computer 
which could simulate the physical behavior of a quantum system as well as perform certain 
tasks much faster than classical computer, attracted immediate attention \cite{preskill,ss}.
The theory of such quantum computers is now well understood and several quantum algorithms like Deutsch-Jozsa (DJ) algorithm,
 Grover's search algorithm, Shor's factorization algorithm, Berstein-Vazirani problem, Hogg's algorithm and quantum counting
 have been developed \cite{deu,grover,shor,hogg,vazi,count1,count2}.

 However, building a realistic large scale quantum computer has been extremely challenging \cite{bou,chuangbook}. Various 
devices are being examined for building a quantum information processing (QIP) device which is coherent and unitary \cite{bou}. 
Among these, nuclear magnetic resonance (NMR) has shown great promise by demonstrating several 
quantum algorithms and other QIP tasks on small-scale devices.\cite{cory97,chuang97,cory98,djchu,djjo,grochu,grojo,ka1,ka,jcp,pram,nat,pram1,ranapra2,ijqi,ranapra1}. 
The last step in any quantum information 
processing task is the ``readout" of the output. Typically in NMR, the readout is obtained by selectively 
detecting spins \cite{jones}, or by mapping out the full density matrix \cite{chutomo,ranabirtomo,newtomo}.  
 It was first pointed out by Ernst et.al. \cite{er} that it is advantageous from the spectroscopic viewpoint  
that quantum states can be assigned to individual spectral lines, 
corresponding to transitions between energy levels  rather than to the energy levels themselves \cite{er}.
However, for such an advantage one has to use an extra qubit called ``observer" qubit.
After computation the readout is obtained by detecting only the observer qubit, 
whose multiplet spectrum provides the result of the computation carried out on the work qubits. 
Such a ``spectral implementation" of a quantum computer was demonstrated by implementation of some logic gates by one- and 
two-dimensional NMR \cite{er}. Later, ``spectral implementation" of a complete set of logic gates and DJ-algorithm \cite{lo}, 
Berstein-Vazirani problem \cite{spec1} and quantum Fourier transform \cite{spec2} has also been implemented by NMR. 
In this work we extend this range by spectrally implementing 
Grover's search algorithm, approximate quantum counting, a modified version of Berstein-Vazirani problem, 
 and Hogg's algorithm. All the algorithms are implemented by both one- and two-dimensional
NMR. To the best of our knowledge this is the first ``spectral implementation" of these algorithms. 

\section{theory}
  A convenient representation of the density matrices of pure states in Liouville space can be obtained by the 
polarization operators for each qubit $(k)$ \cite{er,ernstbook},
\begin{eqnarray}
&&I^k_{0}=\vert 0\rangle \langle0\vert=\pmatrix{1& 0\cr 0& 0} ~~~~~~~~~~~~~~~~~~~
I^k_{1}=\vert 1\rangle \langle1\vert=\pmatrix{0& 0\cr 0& 1} \nonumber \\
&&I^k_{+}=\vert 0\rangle \langle1\vert=\pmatrix{0& 1\cr 0& 0} ~~~~~~~~~~~~~~~~~~~
I^k_{-}=\vert 1\rangle \langle0\vert=\pmatrix{0& 0\cr 1& 0} \nonumber \\
&&I^k_x=\frac{1}{2} (I^k_++I^k_-)=\frac{1}{2}\pmatrix{0& 1\cr 1& 0} ~~~~~~~
I^k_y=\frac{1}{2i} (I^k_+-I^k_-)=\frac{1}{2i}\pmatrix{0& 1\cr -1& 0} \nonumber \\
&&I^k_z=\frac{1}{2} (I^k_0-I^k_1)=\frac{1}{2}\pmatrix{1& 0\cr 0& -1}
\end{eqnarray}  
For example, the density matrix of a pure state $\vert00\rangle+\vert11\rangle$ can be expressed as 
\begin{eqnarray}
\pmatrix{1&0&0&1\cr 0&0&0&0\cr 0&0&0&0\cr 1&0&0&1}&=&
\vert00\rangle\langle00\vert+\vert11\rangle\langle11\vert+\vert00\rangle\langle11\vert+\vert11\rangle\langle00\vert \\
&=&I_{0}^1I_{0}^2+I_{1}^1I_{1}^2+I_{+}^1I_{+}^2+I_{-}^1I_{-}^2
\end{eqnarray}

 The scheme of ``spectral implementation" of one-dimensional (1D) and two-dimensional (2D) NMR is respectively given in figure 1(a) and 1(b). 
We start with the thermal equilibrium density matrix $I^0_z+I^1_z+I^2_z+..+I^N_z$ and in the preparation period 
we create density matrix of the form $I_z^0I^1_0I^2_0..I^N_0$, where
 $I_z^0I^1_0I^2_0..I^N_0=(I^0_0I^1_0I^2_0..I^N_0-I^1_0I^1_0I^2_0..I^N_0)/2$. In this state  
the last N-1 qubits are simultaneously in pseudopure state (PPS) \cite{cory97} in two distinct domains of energy levels,  
in which the observer qubit is in state $\vert 0\rangle$ and $\vert 1\rangle$ respectively. Such a state is 
known as sub-system pseudopure state \cite{chuang97}. This is further elaborated in figure 2.

 The schematic diagram of the energy levels and the spectrum of the observer qubit in a 
three qubit system, where the first qubit is the observer qubit, is given in figure 2. Figure 2(a) shows the equilibrium deviation 
populations (populations in excess of uniform background population) of various energy 
levels and figure 2(b), the equilibrium spectrum of the observer qubit obtained after a $(\pi/2)$ detection pulse. 
Each of the spectral lines in the multiplet, correspond to the state of the other qubits. 
 The energy level diagram along with the deviation populations after creating the desired initial state of 
$I_z^0I_0^1I_0^2$ is given in 2(c). The corresponding observer qubit spectrum has a single line, that of $\vert 00\rangle$,
 indicating that the other qubits are in $\vert 00\rangle$ state. 

 Typically after computation, the density matrix is of the form $I_z^0I^1_{0/1}I^2_{0/1}..I^N_{0/1}$, where the subscript 0/1 means 
that the particular qubit is either in 0 or 1 state.
A subsequent $ (\pi/2)^0_y$ pulse on the observer ($I^0$) qubit 
creates single quantum coherences of the form $I_x^0I^1_{0/1}I^2_{0/1}..I^N_{0/1}$, which gives a 
single line in the spectrum corresponding to the output state of other qubits. An example for the 3-qubit system is given in 
figure 2. Let us assume that we start with the initial $\vert 00\rangle$ pseudopure state of the qubits (other than observer qubit)
and after  some computation let the output state be $\vert 11\rangle$. After such a computation, 
the deviation populations and  spectrum of observer qubit are given respectively in figure 2(e) and 2(f).

 In some algorithms however, the output is a superposition of multiple states. 
Then, the output density  matrix will have non-zero populations 
in all the output states and the coherences between them. 
The spectrum of the observer qubits will thus have multiple lines, 
corresponding to all the output states.  For example, in the 3-qubit system, if the output state 
of the work qubits is $\vert 00\rangle+\vert 11\rangle$, the density matrix is of the form 
$I_z^0I^1_{0}I^2_{0}+I_z^0I_1^{1}I_1^{2}+I_z^0I^1_{+}I^2_{+}+I_z^0I^1_{-}I^2_{-}$. 
After the $(\pi/2)^0_y$ detection pulse on the observer qubit 
the single quantum coherences of the terms $I_x^0I^1_{0}I^2_{0}$ and $I_x^0I^1_{1}I^2_{1}$ will be detected. The spectrum of the 
observer qubit will show two lines corresponding to the states of $\vert 00\rangle$ and $\vert 11\rangle$ of the other qubits. 
The coherences will be converted into multiple quantum coherences which are  not detected directly in NMR. 
Hence, the ``spectral implementation" gives a measure of the deviation populations or probabilities of each state but 
does not measure the coherences, which if required can be measured by state tomography \cite{chutomo,ranabirtomo,newtomo}. 

 A two-dimensional experiment for ``spectral implementation" provides the input and output in the same spectrum. 
 The pulse sequence for the two-dimensional experiment of ``spectral implementation" is given in figure 1(b). 
Suppose a computation starts with the input of $\vert 00..0\rangle$ and end with an output of $\vert 11..1\rangle$ state. 
After preparation of 
the initial $I_z^0I^1_0I^2_0..I^N_0$ state the application of the pulse sequence of figure 1(b) can be analyzed in the following steps:
\begin{eqnarray}
I^0_zI^1_0I^2_0..I^N_0 ~~~~~ &&   \rightarrow{\kern-18pt \raisebox{2pt}{$^{(\pi/2)^0_y}$}}~~~~~  I^0_xI^1_{0}I^2_{0}..I^N_{0} \nonumber \\
                            &&    \rightarrow{\kern-14pt \raisebox{2pt}{$^{t_1}$}}~~~~~~~~~ I^0_xI^1_{0}I^2_{0}..I^N_{0}cos(\omega^0_{00..0}t_1) \nonumber \\
 &&    \rightarrow{\kern-26pt \raisebox{2pt}{$^{(\pi/2)^0_{-y},G_z}$}} ~ I^0_zI^1_{0}I^2_{0}..I^N_{0}cos(\omega^0_{00..0}t_1) \nonumber \\
                            &&   \rightarrow{\kern-14pt \raisebox{2pt}{$^{Comp}$}} ~~~~~   I^0_zI^1_{1}I^2_{1}..I^N_{1}cos(\omega^0_{00..0}t_1) \nonumber \\
            &&  \rightarrow{\kern-24pt \raisebox{2pt}{$^{ (\pi/2)^0_y-t_2}$}} ~~~~~  I^0_xI^1_{1}I^2_{1}..I^N_{1}cos(\omega^0_{00..0}t_1)cos(\omega^0_{11..1}t_2), 
\end{eqnarray}
 where $\omega^0_{00..0}$ and $\omega^0_{11..1}$ are respectively the frequencies of the $\vert 00..0\rangle$ and $\vert 11..1\rangle$ transitions of the 
observer qubit $I^0$, $(\pi/2)^0_y$ is a $(\pi/2)$ rotation of the observer qubit ($I^0$) about y-axis, $G_z$ is the gradient pulse and $Comp$ is the computation 
performed on the work qubits. 
 It may be noted that the signal from the observer qubit is modulated by the frequencies corresponding to both the input and the output states of the work qubits.
 A  series of experiments are performed with systematic increment of the $t_1$ period followed by detection of the observer qubit's signal. 
The collected two-dimensional time domain data set $s(t_1,t_2)$ is double Fourier transformed yielding
a two-dimensional frequency domain spectrum $S(\omega_1,\omega_2)$, which  contains along
$\omega_1$ the input states of work qubits before computation and along $\omega_2$, the output state of work qubits after computation.

\section{Grover's search algorithm}

 Grover's search algorithm can search an unsorted database of size N in $O(\sqrt{N})$ steps while a classical search would require 
$O(N)$ steps \cite{grover}. Grover's search algorithm has been earlier demonstrated by NMR \cite{grochu,grojo}.
The quantum circuit for implementing Grover's search algorithm on two qubit system is given in figure 3(a). 
The algorithm starts from a $\vert 00\rangle$ pseudopure state.  A uniform superposition of all states are created by 
the initial Hadamard gates $(H)$. Then the sign of the searched state $``x"$ is inverted by the oracle through the operator 
\begin{eqnarray}
U_x=I-2\vert x\rangle \langle x \vert.  
\end{eqnarray}
 An inversion about mean is performed on all the states by a diffusion operator $HU_{00}H$, where 
\begin{eqnarray}
U_{00}=I-2\vert 00\rangle \langle 00 \vert.
\end{eqnarray}
 For an N-sized database the algorithm requires $O(\sqrt{N})$ iterations of $U_x HU_{00}H$. 
For a 2-qubit system with four states, only one iteration is required. 
We have implemented this algorithm on the two qubits of a three qubit system with the third qubit 
acting as the observer qubit. The three qubit system chosen for this purpose is 
4-fluro 7-nitro benzofuran (dissolved in CDCl$_3$), which comprises of a two protons ($^1$H) and a flourine ($^{19}$F).
The chemical structure of the molecule along with the equilibrium proton and fluorine spectrum is given in figure 4(a). 
We have chosen the fluorine spin as the observer qubit. The Hamiltonian of the system is
\begin{eqnarray}
{\mathcal H}=\sum_{i=0}^{2} 2\pi\nu_i I^i_z +\sum_{i>j} 2\pi J_{ij} I^i_zI^j_z,
\end{eqnarray}
where $\nu_i$ are the resonance frequencies of various spins and $J_{ij}$ are the indirect couplings. 
The experiments were performed at a field of 11.4 Tesla in a Bruker DRX500 spectrometer.
At the magnetic field of 11.4 Tesla, the resonant frequency of proton is 500.13 MHz and that of fluorine is 470.59 MHz.
The frequency difference between the two protons is 646 Hz.
The J-couplings are $J_{01}$= -3.84 Hz, $J_{02}$=8.01 Hz and
$J_{12}$= 8.07 Hz. The $^1$H transmitter frequency is set at the center of the proton spectrum.

The required initial state of $I_z^0 \vert 00\rangle \langle 00\vert$ was prepared by the method of 
pair of pseudopure states (POPS), originally suggested by Fung \cite{fung00,fungjcp}. The method requires two population 
distributions, (i) equilibrium populations and (ii) population distribution after a selective $(\pi)$-pulse 
on $\vert 000\rangle \leftrightarrow \vert 100\rangle$. Subtraction of (ii) from (i) effectively gives the 
initial state of $I_z^0 \vert 00\rangle \langle 00\vert$ (figure 4(b) corresponding to the schematic PPS of figure 2(c)). 
It might be noted that the method of creation of sub-system pseudopure states from cat-states can also be 
adopted for creation of this initial state \cite{cat}.

 The Hadamard gates are implemented by $(\pi/2)^{1,2}_{-y} (\pi)^{1,2}_x $ pulse (pulses are applied from 
left to right) \cite{grochu}, where $(\theta)^{1,2}_{x}$ 
denotes a $\theta$-angle pulse (rotation) on 1$^{st}$ and 2$^{nd}$-qubit about the $x$-axis.  
The $U_{00}$ operator is a controlled phase gate which can be implemented by the sequence 
$[(\tau/2)(\pi)^{1,2}_x(\tau/2)(\pi)^{1,2}_x][(\pi/2)^{1,2}_{-y}(\pi/2)^{1,2}_{-x}(\pi/2)^{1,2}_{y}]$, where 
$\tau=1/2J_{12}$ \cite{grochu}. The sequence $[(\tau/2)(\pi)^{1,2}_x(\tau/2)(\pi)^{1,2}_x]$ evolves the system only under 
the J$_{12}$-coupling and refocuses all other couplings and proton chemical shifts \cite{ernstbook}, whereas the 
$[(\pi/2)^{1,2}_{-y}(\pi/2)^{1,2}_{-x}(\pi/2)^{1,2}_{y}]$ is a composite z-rotation on both the qubits \cite{levitt}.
Similarly, the other phase gates can be constructed as \cite{grochu},
\begin{eqnarray}
&&U_{01}=[(\tau/2)(\pi)^{1,2}_x(\tau/2)(\pi)^{1,2}_x][(\pi/2)^{1,2}_{-y}(\pi/2)^{1}_{x}(\pi/2)^{2}_{-x}(\pi/2)^{1,2}_{y}] \nonumber \\
&&U_{10}=[(\tau/2)(\pi)^{1,2}_x(\tau/2)(\pi)^{1,2}_x][(\pi/2)^{1,2}_{-y}(\pi/2)^{1}_{-x}(\pi/2)^{2}_{x}(\pi/2)^{1,2}_{y}] \nonumber \\
&&U_{11}=[(\tau/2)(\pi)^{1,2}_x(\tau/2)(\pi)^{1,2}_x][(\pi/2)^{1,2}_{-y}(\pi/2)^{1,2}_{x}(\pi/2)^{1,2}_{y}] 
\end{eqnarray}
 The pulses which are simultaneously applied on both the qubits are achieved by hard pulses. 
However, some gates require selective excitation of qubits.
 Since the resonance frequencies of the two protons are relatively close to each other, selective excitation 
of a particular proton qubit requires long low-power pulses, which introduce significant errors in the computation \cite{djjo,grojo}.       
Fortunately, in case there are two homonuclear qubits, the selective pulses can be substituted by hard pulses and 
delays using the variation of ``jump-and-return'' sequence  \cite{jump}, as demonstrated by Jones et.al. \cite{nmrcount}. 
For example, the pulse sequence of $U_{01}$ gate 
requires $(\pi/2)^{1}_{x}(\pi/2)^{2}_{-x}$ at one point. This can be achieved by using the identity \cite{levitt}
\begin{eqnarray}
(\pi/2)_{-y}(\pi/2)_{\pm z}(\pi/2)_{y}=(\pi/2)_{\pm x}. 
\end{eqnarray}
If the proton transmitter frequency is set at the center of the spectrum, then $\nu_1=-\nu_2=\nu$, and,  a delay of $(1/4\nu)$
evolves the two protons under the Zeeman Hamiltonian of $2\pi\nu(I_z^1-I_z^2)$ to give the intermediate $(\pi/2)_{\pm z}$ 
rotation of Eq.[9]. Hence, 
\begin{eqnarray}
(\pi/2)^{1}_{x}(\pi/2)^{2}_{-x}=(\pi/2)_{-y}-(1/4\nu)-(\pi/2)_{y}.
\end{eqnarray}
 Similarly, the pulse $(\pi/2)^{1}_{-x}(\pi/2)^{2}_{x}$ required for $U_{10}$ gate,
 can be achieved by 
\begin{eqnarray}
(\pi/2)^{1}_{-x}(\pi/2)^{2}_{x}=(\pi/2)_{y}-(1/4\nu)-(\pi/2)_{-y}     
\end{eqnarray}

 In principle however, the evolution under J-coupling during $(1/4\nu)$ would lead to some non-ideal characteristics \cite{nmrcloning}, 
which is 
minimal in our system, since the ratio of maximum J-coupling to chemical shift frequency difference $\sim$ 1:80.
This error is significantly less than the error introduced due to evolution under internal Hamiltonian during low 
power long duration qubit selective pulses.

  After application of the quantum circuit in figure 3(a) on the initial state of $I^z_0 \vert 00\rangle \langle 00\vert$, 
the observer qubit was detected by a $(\pi/2)$ pulse. From the obtained spectrum given in figure 5(a), 
one can identify the searched state ($\vert x\rangle$) directly. 
The two-dimensional experiment for ``spectral implementation" has the added advantage that the input and output can be 
identified in a single spectrum. The 2D experiment of figure 1(b) was carried out, where during the computation period, the quantum circuit 
of figure 3, was implemented on $I_1$ and $I_2$. The resultant spectrum given in figure 5(b), shows the input and output in each case. 
For example, when $\vert x\rangle=\vert 11\rangle$, a cross-peak at the frequency of $\vert 00 \rangle $ 
transition along $\omega_1$ to that of the $\vert 11 \rangle $ transition along $\omega_2$, identifies 
 the input as $\vert 00\rangle$ and the output as $\vert 11\rangle$. The 2D spectra in figure 5(b) contains the initial state 
of $\vert 00\rangle$ and the searched state of $\vert 00\rangle$, $\vert 01\rangle$, $\vert 10\rangle$ and $\vert 11\rangle$.
In the 1D spectrum of Fig. 5(a), the intial PPS state has to be ascertained independently prior to the implementation of 
search algorithm.
 
\section{approximate quantum counting}
The search problem may be thought as finding $k$-entries out of $N$, which satisfy the  condition $f(x)=1$. For the other, 
$N-k$ entries, $f(x)=0$. While Grover's search algorithm searches these k-items (one at a time), quantum counting 
finds out the value of $k$ \cite{count1,count2}. This has extreme importance because in case of multiple solutions, 
the required number of Grover's iteration 
scales as $O(\sqrt{N/k})$ \cite{chuangbook}. 
Hence finding out the number of existing solutions speeds up the search procedure. Moreover, 
the fact that counting can find out whether the number of solutions is zero or finite,
makes it applicable to the non-deterministic (NP)-complete search problems, where it is important to know 
whether solution exists for a given search problem \cite{chuangbook}.   
Approximate quantum counting has been demonstrated using NMR by Jones and Mosca \cite{nmrcount}. In this work we provide a 
``spectral implementation" of approximate quantum counting in the three qubit system of 4-fluro 7-nitro benzofuran, where 
the $I^2$ is the target qubit, $I^1$ the control qubit and the $I^0$, the observed qubit.  

The working of counting algorithm, as detailed by Jones and Mosca \cite{nmrcount}, is as follows.
 Counting algorithm can be thought of as 
a method for estimating the eigenvalue of Grover's iteration $G= HU_0H^{-1}U_f$, where 
$U_0=I-2\vert 00..0\rangle \langle 00..0\vert$ and $U_f$ transforms 
$\vert x\rangle$ to $(-1)^{f(x)+1}\vert x\rangle$. 
Starting from the initial $\vert 00..0\rangle$ state, an initial Hadamard on target qubit creates 
an uniform superposition $H\vert 00..0 \rangle$=$(\vert \psi_+\rangle +\vert \psi_-\rangle)/\sqrt2$, where
 $\vert \psi_+ \rangle$ and $\vert \psi_- \rangle$ are two eigenvectors of G \cite{nmrcount}. 
These two eigenvectors are with eigenvalues of $e^{\pm i\phi_k}$, where $sin(\phi_k/2)=\sqrt{k/N}$.  
An uniform superposition of the control qubit is also created. The application of controlled $G$ produces 
the result  
\begin{eqnarray}
\vert \psi^1_+\rangle=(\vert 0\rangle +e^{i\phi_k}\vert 1\rangle)\vert\psi_+\rangle/\sqrt(2).
\end{eqnarray} 
If $r$ iterations are performed, then the state is 
\begin{eqnarray}
\vert \psi^r_+\rangle=(\vert 0\rangle +e^{ir\phi_k}\vert 1\rangle)\vert\psi_+\rangle/\sqrt(2).
\end{eqnarray}

A second Hadamard gate on the control qubit produces
\begin{eqnarray}
\vert \psi^r_{f+}\rangle= [(1+e^{ir\phi_k})\vert 0\rangle +(1-e^{ir\phi_k})\vert 1\rangle]\vert\psi_+\rangle/2.
\end{eqnarray}
 A similar result will happen in the case of $\vert \psi_-\rangle$. At the end, the final state $\vert \psi_f\rangle$ will be an 
entangled state of the control and target qubits, except when $k=0$ or $k=N$ \cite{count1,count2,nmrcount}.
 
  Jones et.al. have implemented the quantum circuit of figure 3(b) in a two-qubit system, 
measured the signal from control qubit, thereby tracing the target qubit, and shown that the signal
 assumes a sinusoidal behavior with $r$ whose frequency depend on $\phi_k$ \cite{nmrcount}. 
We have instead, started from the initial $I_z^0 \vert 00\rangle \langle 00\vert$ state 
and inferred the result of counting from the spectrum of observer qubit. For a two qubit 
case only one Grover's iteration is sufficient to get the result \cite{nmrcount}.
Given in Table 1 are the count $k$, their 
corresponding $\phi$, the $U_f$  operators and final state of the system for $r=1$. Note that for $k=0$, the final state is 
$\vert \psi_f \rangle=\vert 00\rangle$ and for $k=N=2$ the final state is $\vert \psi_f \rangle=\vert 10\rangle$.  
For $k=1$ the output states are in entangled form of all the states, $\vert 00\rangle$, $\vert 01\rangle$, $\vert 10\rangle$ and $\vert 11\rangle$.

 Starting with the initial state of $I_z^0 \vert 00\rangle \langle 00\vert$, we implemented the quantum circuit 
of figure 4(b). The controlled controlled $U_0$ and controlled $U_{f_{01}}$ have the same operator as that of 
two-qubit controlled phase gate $U_{10}$ implemented in Grover's algorithm (section II), whose corresponding pulse sequence is  
given in Eq. [11]. 
$U_{f_{10}}$ has the same operator and pulse sequence as that of $U_{11}$ in Eq. [11].  
$U_{f_{11}}$ is an identity operator and required no pulses. $U_{f_{00}}$ required a $(\pi)^1_z$ rotation.
This $(\pi)^1_z$ rotation was implemented with hard pulses and evolution under Zeeman Hamiltonian, 
\begin{eqnarray}
(\pi)^1_z=(1/4\nu)-[(\pi/2)^{1,2}_{-y}(\pi/2)^{1,2}_{x}(\pi/2)^{1,2}_{y}]. 
\end{eqnarray}
During the delay $(1/4\nu)$ the system
evolves under the Zeeman Hamiltonian to acquire a rotation of $(\pi/2)^1_{z}(\pi/2)^2_{-z}$. The subsequent composite
z-pulse was applied on both qubits, $(\pi/2)^{1,2}_{z}=[(\pi/2)^{1,2}_{-y}(\pi/2)^{1,2}_{x}(\pi/2)^{1,2}_{y}]$,
which cancels the rotation of second qubit but adds to the rotation of first qubit to give an effective $(\pi/2)^1_z$ rotation.
It may be noticed that there are two pseudo-Hadamard gates on second qubit which require spin-selective pulses 
since h=$(\pi/2)_y$ and h$^{-1}$=$(\pi/2)_{-y}$. However, these pulses can also  be performed by hard pulses and 
evolution under Zeeman Hamiltonian using the "jump-and-return" logic\cite{nmrcount}. 
\begin{eqnarray}
(\pi/2)^{2}_y=(\pi/2)^{1,2}_{x}-(1/8\nu)-(\pi/2)^{1,2}_{-x}(\pi/4)^{1,2}_{y}, \nonumber \\ 
(\pi/2)^{2}_{-y}=(\pi/2)^{1,2}_{-x}-(1/8\nu)-(\pi/2)^{1,2}_{x}(\pi/4)^{1,2}_{-y}. 
\end{eqnarray}
After implementing the quantum circuit of figure 3(b), the observer qubit was measured. 
The observer qubit's spectrum given in figure 6(a), shows four lines for 
 $k=1~(f_{01} ~\mathrm{and}~ f_{10})$. For $k=0~(f_{00})$, only $\vert 00\rangle$ transition and for $k=2~(f_{11})$, 
only $\vert 10\rangle$ transition is observed. 
The 2D-spectrum of figure 6(b) contains correlation of the output state with the initial  $\vert 00\rangle$ 
pseudopure state, and confirms the same result.

\section{Berstein-Vazirani problem}

   Berstein and Vazirani considered the problem of determining a n-bit string $``a"$ \cite{vazi}. Classically each query would yield 
one bit of information and hence would  require n-queries to the database. However, Berstein and Vazirani showed that 
a quantum algorithm can solve the problem with one quantum query \cite{vazi}. For this purpose, the oracle has to compute a function 
$f_a(x)=a.x$ . The scheme proposed by Berstein and Vazirani required an ancillary qubit and determined a n-qubit string 
with n+1 qubits, which has been demonstrated by NMR recently \cite{spec2}. However, Du and his co-workers had simplified the scheme 
such that the ancillary qubit was not required \cite{nmrvazi}. We have implemented the  Du-scheme, since it has the advantage of 
determining a n-qubit string with n-qubit system. The quantum circuit of a two-qubit implementation is given in figure 3(c). 
Starting from $\vert 0\rangle^n$, the Hadamard gates create an uniform superposition
\begin{eqnarray}
 \vert \psi_1 \rangle=\frac{1}{2^{n/2}} \sum_{x=0}^{2^n-1}\vert x\rangle    
\end{eqnarray}
 The $U_a$ operator transforms $\vert x\rangle \rightarrow (-1)^{f_a(x)} \vert x\rangle$. 
The unitary operator $U_a$ can be decomposed into direct products of single-qubit operations \cite{nmrvazi}
\begin{eqnarray}
U_a&=&U^1\otimes U^2\otimes...\otimes U^n, \nonumber \\
U^i&=&I, ~~~~~ a_i=0  \nonumber \\
   &=&\sigma_z,~~~~ a_i=1 \nonumber \\
I&=&\pmatrix{1 & 0 \cr 0 & 1},~~~~~ \sigma_z=\pmatrix{1 & 0 \cr 0 & -1} 
\end{eqnarray}
Operation of $U_a$ creates a new state $\vert \psi_2 \rangle$ of the form, 

\begin{eqnarray}
 \vert \psi_2 \rangle=U_a \vert \psi_1 \rangle=\frac{1}{2^{n/2}} \sum_{x=0}^{2^n-1} (-1)^{a.x} \vert x\rangle.
\end{eqnarray}
 The final state after the subsequent Hadamard operation is 
\begin{eqnarray}
 \vert \psi_3 \rangle=H \vert \psi_2 \rangle=\frac{1}{2^{n}} \sum_{x=0}^{2^n-1} \sum_{x=0}^{2^n-1} (-1)^{a.x} (-1)^{x.y} \vert y\rangle
\end{eqnarray}
 However, since
$\sum_{x=0}^{2^n-1} (-1)^{a.x} (-1)^{x.y}=\delta_{a,y}$ \cite{preskill}, $\vert \psi_3 \rangle = \vert a \rangle$ \cite{vazi,nmrcount}.
    
 The algorithm was implemented to determine a two-qubit string by ``spectral implementation" using three 
qubit system of 4-fluro 7-nitro benzofuran. After creating POPS, followed by Hadamard pulses, the operator $U_a$ 
was applied for $\vert a\rangle= \vert 00\rangle$, $\vert 01\rangle$, $\vert 10\rangle$  and $\vert 11\rangle$.    
$U_{00}$ is unity operator and does not require any pulse. $U_{10}$ is $\sigma^1_z$, which requires a $(\pi)^1_z$ rotation. 
Once again, the $(\pi)^1_z$ rotation was implemented using the pulse sequence of Eq.[15].
Similarly, $U_{01}$ was implemented by 
\begin{eqnarray}
(\pi)^2_z=(\pi)^{1,2}_{x}(1/4\nu)(\pi)^{1,2}_{x}-[(\pi/2)^{1,2}_{-y}(\pi/2)^{1,2}_{x}(\pi/2)^{1,2}_{y}]. 
\end{eqnarray}

$U_{11}$ is $\sigma^1_z\sigma^2_z$, which can be achieved by a composite z-pulse of 
$(\pi)^{1,2}_z=(\pi/2)^{1,2}_{-y}(\pi)^{1,2}_{x}(\pi/2)^{1,2}_{y}$. After application of the final Hadamard 
pulses, the observer qubit was detection by a $(\pi/2)$ pulse. The obtained spectrum given in figure 7(a), clearly 
determines the 2-bit string in each case. The result of 2D experiment is given in figure 7(b). The 2D spectrum correlates the 
input $\vert 00\rangle$ to the output in each case.       

    The above algorithm was also implemented to determine a three-qubit string by ``spectral implementation" using a 
four-qubit system. 
  The molecule 2-3 diflouro 6-nitrophenol (dissolved in CDCl$_3$+1 drop D$_2$O) has 4 weakly coupled spin-1/2
nuclei. The proton of the phenol group is exchanged with the D$_2$O.
 The two remaining protons and the two fluorine nuclei constitute the four-qubit system.
 The equilibrium spectrum of each nucleus  is given in Fig. 8 (a). In a 500 MHz NMR spectrometer,
the chemical shift difference between the two Fluorine spins is 16 kHz while that between the two  protons is 560 Hz. 
The couplings range from 19.13 Hz to -2.4 Hz. 

 The operators and pulse sequences required for each string of 
a three qubit system is given in Table 2. Since the chemical shift difference between the two fluorine 
spins are considerably large (16kHz), selective pulses do not introduce significant errors. 
 The pulses on fluorine spin $I^3$ were achieved 
by Gaussian shaped selective pulses of 12.5$\mu$s duration. The proton transmitter frequency 
was kept at the middle of the spectrum and the selective z-pulses on protons were 
applied in similar logic as in the two-qubit case (Eq. [15] and [21]). Hard pulses were applied when both protons had to be  pulsed 
simultaneously. The algorithm was implemented starting from the 
initial state of $I^0_z\vert 000\rangle \langle 000\vert$ (figure 8(b)) and finally the observed qubit was measured 
by selective Gaussian shaped $(\pi/2)_y$ pulse. The only transition present in each spectrum given in figure 9(a)
indicates the corresponding string. The 2D experimental spectra given in figure 9(b) verify the same results, correlating 
the input state of $\vert 000\rangle$ in each case.

\section{Hogg's algorithm}
 Satisfiability (SAT) problem is one of the nondeterministic polynomial (NP) combinatorial search problems \cite{hogg}. 
SAT problem consists of a logical formula 
in $n$ variables, $V_1,V_2,...,V_n$ \cite{hogg}. One has to find an assigment (true or false) for each variable $V_i$, such that 
it makes the formula true. The logical formula can be expressed in various equivalent forms, as 
conjunction of clauses, where a clause in a disjunction of some variables. 
A clause with $k$ variables is false for exactly one set of values for its variables but true for the 
other $2^k-1$ sets. An example of clause for k=3 is $V_1$ OR $V_2$ OR $V_3$, where the clause is $false$ 
for only $V_1=V_2=V_3=false$. Only the assignments which satisfy all the clauses are considered as solutions \cite{hogg}.

 While the number of steps required by a classical algorithm increase linearly with the size of the variables \cite{hogg},
 Hogg's algorithm can solve 1-SAT and maximally constrained $k$-SAT problems in a single step, whatever be 
 the size of the variable \cite{hogg}. Hogg's algorithm has been succesfully implemented by NMR in a three qubit system \cite{nmrhogg}. 
 Here we demonstrate Hogg's algorithm by ``spectral implementation".
The Hogg's algorithm starts by creating a uniform superposition of states by 
 initializing from $\vert \psi_0\rangle=\vert 0\rangle^{\otimes n}$ and applying Hadamard gate on all 
qubits $\vert \psi_1\rangle=H^{\otimes n}\vert \psi_0\rangle=2^{-n/2}\sum_{s} \vert s \rangle$. Let $m$ be the 
number of clauses in the 1-SAT logical formula. Then the unitary operations $UR$ are consecutively applied 
to yield the final state $\vert \psi_f\rangle=UR\vert \psi_1\rangle$ where U and R defined as follows. R 
adjusts the phases of $\vert s\rangle$ depending on the conflicts $c$ of the different assignments in the 
superposition of $s$, ranging from 0 to $m$. R is a diagonal matrix of the form
\begin{eqnarray}
R^{ss}&=&\sqrt{2} cos[(2c-1)\pi/4], ~~~~~~~\mathrm{for ~~even~~} m  \nonumber \\
   &=&i^c ~~~~~~~~~~~~~~~~~~~~~~~~~~~~\mathrm{for~~ odd~~}m
\end{eqnarray}
      The operator U mixes the amplitudes from different assignments with elements $U_{rs}$, depending the 
Hamming distance $d$ between $r$ and $s$. U is of the form
\begin{eqnarray}
U_{rs}&=&U_{d(r,s)}  \nonumber \\
&=&2^{-(n-1)/2}cos[(n-m+1-2d)\pi/4]  ~~~~~~~\mathrm{for ~~even~~} m  \nonumber \\
&=&2^{-n/2}e^{i\pi(n-m)/4}(-i)^d ~~~~~~~~~~~~~~~~~~~~~~~~\mathrm{for~~ odd~~}m 
\end{eqnarray}
where $d=\vert r\vert +\vert s\vert-\vert r\wedge s\vert$ is the Hamming distance between $r$ and $s$, i.e. 
number of positions at which their values differ. $U$ can be decomposed into $H\Gamma H$ where H is the 
Hadamard matrix and $\Gamma$ is a diagonal matrix of the form 
\begin{eqnarray}
\Gamma_{rr}=\gamma(r)=\gamma_h&=&\sqrt{2}cos[(m-2h-1)\pi/4] ~~~~~~~\mathrm{for ~~even~~} m \nonumber \\
&=& i^he^{-i\pi m/4} ~~~~~~~~~~~~~~~~~~~~~~~~~~~~\mathrm{for~~ odd~~}m
\end{eqnarray}
where $h=\vert r\vert$, and hence $\Gamma_{rr}$ depend on the number of 1-bits in each assignment. For a 
detailed description of the working of the algorithm see ref \cite{hogg}. Hence the Hogg's quantum
starts with the initial $\vert 00\rangle$ state and reaches the desired output state $\vert \psi_f\rangle$ by
\begin{eqnarray}
\vert \psi_f\rangle=URH\vert00\rangle
\end{eqnarray}  
and the final step is to measure the output state $\vert \psi_f\rangle$. 
We have observed the final state through detection of the observer qubit.
 
 The quantum circuit of Hogg's algorithm for a three-qubit system is given in figure 3(d).
While implementing the corresponding pulse sequence, the consecutive pulses of opposite phases cancel out,
yielding a simplified sequence \cite{nmrhogg}. The $m=1$ and 3 clauses, 
their logic formulae and the reduced pulse sequences are given in table 3 \cite{nmrhogg}. 
Only the $m$=1 and 3 cases are demonstrated here for ease of implementation.
The selective pulses on fluorine spin were achieved using  
Gaussian shaped pulses. In protons, the selective pulses 
 were achieved by hard pulses concatenated with Zeeman evolution (Eq. [10], [11] and [16]). 
 For example, while implementing $V_1$,
\begin{eqnarray}
V_1:[(\pi)^1_x][(\pi/2)^2_y][(\pi/2)^3_y]&=&[(\pi/2)^{1,2}_{-y}-1/4\nu-(\pi/2)^{1,2}_{y}(\pi/2)^{1,2}_{x}] \nonumber \\
&&[(\pi/2)^{1,2}_{-x}-1/8\nu-(\pi/2)^{1,2}_{x}(\pi/4)^{1,2}_{y}][(\pi/2)^3_y] \nonumber \\
&=& [(\pi/2)^{1,2}_{-y}-1/4\nu-(\pi/2)^{1,2}_{y}-1/8\nu-(\pi/2)^{1,2}_{x}(\pi/4)^{1,2}_{y}][(\pi/2)^3_y] 
\end{eqnarray} 
The spectra obtained by the 
one-dimensional experiment is given in figure 10 while the spectra obtained in 2D-experiment is given in figure 11. 
The spectra of observer qubit clearly identifies the desired outputs of table 3.  
For example, in the case of $V_1$, the output has all the states that satisfy the condition that 
1$^{st}$ qubit is $\vert 1\rangle$ or  $false$; namely $\vert 001\rangle$, $\vert 011\rangle$, 
$\vert 101\rangle$ and $\vert 111\rangle$ (read the order of qubits from right to left). 
Similarly, for $V_3\wedge V_2 \wedge V_1$, the output consists of 
the sole answer $\vert 111\rangle$, which satisfies the condition that all the qubits are in state $\vert 1\rangle$.  

\section{conclusion}
 We have demonstrated ``spectral implementation" of several quantum algorithms by one- and two-dimensional NMR.
Provided future quantum computers run with high fidelity, ``spectral implementation" delivers an aphoristic method of 
readout. Though it requires the use of an observer qubit, this qubit also helps in creating a pseudopure state 
by non-scalable and effective methods like POPS \cite{fung00}. With the essentiality that the observer qubit has resolved 
spectrum, the principle of ``spectral implementation" is applicable to higher qubit systems without increasing complexity.    
 
\section{acknowledgment}
Useful discussions with Prof. G. S. Agarwal are gratefully acknowledged. 
The use of DRX-500 NMR spectrometer funded by the Department of
Science and Technology (DST), New Delhi, at the Sophisticated Instruments Facility, 
Indian Institute of Science, Bangalore, is also gratefully acknowledged. 
AK acknowledges "DAE-BRNS" for the award of "Senior Scientists scheme" and DST for a research grant for
 "Quantum Computing by NMR".

%=====================================================================================

\pagebreak
Table 1 :The various possible the count $k$ for a two-qubit system, their corresponding $\phi$, the $U_f$  operators and final 
state of the system $\vert \psi_{output}\rangle$.\\
\begin{tabular}{|c|c|c|c|} \hline \hline
~~~k~~~~&~~~~$\phi_k$~~~~&~~~~$U_f$~~~~&~~~~$\vert \psi_{output}\rangle$~~~~ \\ \hline
~~~0~~~~&~~~~0~~~~&~~~~$U_{f_{00}}= \pmatrix{1 &0& 0& 0& \cr 0& 1& 0& 0\cr 0& 0& -1& 0\cr 0& 0& 0& -1}$~~~~&~~~~$\vert 00 \rangle$~~~~ \\
~~~1~~~~&~~~~$\pi/2$~~~~&~~~~$U_{f_{01}}= \pmatrix{1 &0& 0& 0& \cr 0& 1& 0& 0\cr 0& 0& -1& 0\cr 0& 0& 0& 1}$
~~~~&~~~~$\vert 00 \rangle-\vert 01 \rangle+\vert 10 \rangle+\vert 11 \rangle$~~~~ \\
~~~1~~~~&~~~~$\pi/2$~~~~&~~~~$U_{f_{10}}= \pmatrix{1 &0& 0& 0& \cr 0& 1& 0& 0\cr 0& 0& 1& 0\cr 0& 0& 0& -1}$
~~~~&~~~~$\vert 00 \rangle+\vert 01 \rangle+\vert 10 \rangle-\vert 11 \rangle$~~~~ \\
~~~2~~~~&~~~~$\pi$~~~~&~~~~$U_{f_{11}}= \pmatrix{1 &0& 0& 0& \cr 0& 1& 0& 0\cr 0& 0& 1& 0\cr 0& 0& 0& 1}$~~~~&~~~~$\vert 10 \rangle$~~~~\\
\hline
\end{tabular}

\pagebreak
Table 2: The operators and pulse sequences required for determination of each string in 
a three qubit system \\
\hspace*{5cm}
\begin{tabular}{|c|c|c|} \hline \hline
~~~string~~~~&~~~~Operator~~~~&~~~~Pulse sequence~~~~ \\ \hline
~~~$\vert000\rangle$~~~~&~~~~I~~~~&~~~~no pulse~~~~ \\
~~~$\vert001\rangle$~~~~&~~~~$\sigma^3_z$~~~~&~~~~$(\pi)_z^3$~~~~ \\
~~~$\vert010\rangle$~~~~&~~~~$\sigma^2_z$~~~~&~~~~$(\pi)_z^2$~~~~ \\
~~~$\vert011\rangle$~~~~&~~~~$\sigma^2_z\sigma^3_z$~~~~&~~~~$(\pi)_z^2(\pi)_z^3$~~~~\\
~~~$\vert100\rangle$~~~~&~~~~$\sigma^1_z$~~~~&~~~~$(\pi)_z^1$~~~~ \\
~~~$\vert101\rangle$~~~~&~~~~$\sigma^1_z\sigma^3_z$~~~~&~~~~$(\pi)_z^1(\pi)_z^3$~~~~ \\
~~~$\vert110\rangle$~~~~&~~~~$\sigma^1_z\sigma^2_z$~~~~&~~~~$(\pi)_z^{1,2}$~~~~ \\
~~~$\vert111\rangle$~~~~&~~~~$\sigma^1_z\sigma^2_z\sigma^3_z$~~~~&~~~~$(\pi)_z^{1,2}(\pi)_z^{3}$~~~~ \\
\hline
\end{tabular}

\pagebreak
Table 3: Logic formulae for m=1 or m=3 in a 3-qubit system, corresponding pulse sequences, and theoritical solutions \cite{nmrhogg}. 
Read the order of qubits from right to left.  \\

\begin{tabular}{|c|c|c|c|} \hline \hline
m~~&~~~~Logic formula~~~~&~~~~Reduced pulse sequence~~~~&~~~~Final state $\vert \psi_f\rangle$~~~~ \\ \hline
~&~~~$V_1$~~~~&$(\pi)^1_x(\pi/2)^2_y(\pi/2)^3_y$&$\vert 001\rangle+\vert 011\rangle+\vert 101\rangle+\vert 111\rangle$\\
~&~~~$\bar V_1$~~~~&$(\pi/2)^2_y(\pi/2)^3_y$&$\vert 000\rangle+\vert 010\rangle+\vert 100\rangle+\vert 110\rangle$ \\

1&~~~$V_2$~~~~&~~~~$(\pi/2)^1_y(\pi)^2_x(\pi/2)^3_y$&$\vert 010\rangle+\vert 011\rangle+\vert 110\rangle+\vert 111\rangle$ \\
~&~~~$\bar V_2$~~~~&~~~~$(\pi/2)^1_y(\pi/2)^3_y$&$\vert 000\rangle+\vert 001\rangle+\vert 100\rangle+\vert 101\rangle$ \\
~&~~~$V_3$~~~~&~~~~$(\pi/2)^1_y(\pi/2)^2_y(\pi)^3_x$&$\vert 100\rangle+\vert 101\rangle+\vert 110\rangle+\vert 111\rangle$ \\
~&~~~$\bar V_3$~~~~&~~~~$(\pi/2)^1_y(\pi/2)^2_y$&$\vert 000\rangle+\vert 001\rangle+\vert 010\rangle+\vert 011\rangle$ \\ \hline
~&~~~$V_3\wedge V_2 \wedge V_1$~~~~&~~~~$(\pi/2)^{1,2,3}_x(\pi/2)^{1,2,3}_{-y}(\pi/2)^{1,2,3}_x$&$\vert 111\rangle$ \\
~&~~~$V_3\wedge V_2 \wedge \bar V_1$~~~~&~~~~$(\pi/2)^{1}_{-x}(\pi/2)^{2,3}_x(\pi/2)^{1,2,3}_{-y}(\pi/2)^{1,2,3}_x$&$\vert 110\rangle$ \\
~&~~~$V_3\wedge \bar V_2 \wedge V_1$~~~~&~~~~$(\pi/2)^{1,3}_x(\pi/2)^{2}_{-x}(\pi/2)^{1,2,3}_{-y}(\pi/2)^{1,2,3}_x$&$\vert 101\rangle$ \\
3&~~~$V_3\wedge \bar V_2 \wedge \bar V_1$~~~~&~~~~$(\pi/2)^{1,2}_{-x}(\pi/2)^{3}_x(\pi/2)^{1,2,3}_{-y}(\pi/2)^{1,2,3}_x$&$\vert 100\rangle$ \\
~&~~~$\bar V_3\wedge V_2 \wedge V_1$~~~~&~~~~$(\pi/2)^{1,2}_x(\pi/2)^{3}_{-x}(\pi/2)^{1,2,3}_{-y}(\pi/2)^{1,2,3}_x$&$\vert 011\rangle$ \\
~&~~~$\bar V_3\wedge V_2 \wedge \bar V_1$~~~~&~~~~$(\pi/2)^{1,3}_{-x}(\pi/2)^{2}_x(\pi/2)^{1,2,3}_{-y}(\pi/2)^{1,2,3}_x
$&$\vert 010\rangle$ \\
~&~~~$\bar V_3\wedge \bar V_2 \wedge V_1$~~~~&~~~~$(\pi/2)^{1}_x(\pi/2)^{2,3}_{-x}(\pi/2)^{1,2,3}_{-y}(\pi/2)^{1,2,3}_x
$&$\vert 001\rangle$ \\
~&~~~$\bar V_3\wedge \bar V_2 \wedge \bar V_1$~~~~&~~~~$(\pi/2)^{1,2,3}_{-x}(\pi/2)^{1,2,3}_{-y}(\pi/2)^{1,2,3}_x
$&$\vert 000\rangle$ \\
\hline
\end{tabular}

\pagebreak
{\Large Figure Captions}

 Figure 1. Experimental protocol for ``spectral implementation" of quantum algorothms \cite{er}. (a) One-dimensional 
experiment. The first stage is to create an subsystem pseudopure state of the type $I_z^0I^1_0I^2_0..I^N_0$, followed 
by computation on $I^1..I^N$ qubits. Finally the transitions of the observer qubit $I^0$ are detected by a $90^o_y$ pulse.
(b) Two-dimensional experiment. After the creation of initial $I_z^0I^1_0I^2_0..I^N_0$ subsystem PPS, the 
observer qubit is flipped by $90^o_y$ pulse to transverse magnetization and allowed to evolve for a time $t_1$. 
During $t_1$, the transitions of the observer qubit modulate with the frequencies characterized by the input state of the other N qubits.
A subsequent $90^o_{-y}$ brings the magnetization back to longitudinal direction. The computation is performed 
on the $I^1..I^N$ qubits. The transitions of the observer qubit are finally detected by a $90^o_y$ pulse.  
A  series of experiments are performed with systematic increment of the $t_1$ period and the collected 2D data set $s(t_1,t_2)$ 
 is Fourier transformed to get the 2D spectrum $S(\omega_1,\omega_2)$.

 Figure 2. Schematic diagram of the energy levels and the spectrum of the observer qubit at different stages of ``spectral implementation".
(a) Deviation equlibrium populations. The dotted arrows denote the transitions of observer qubit.
(b) Equilibrium spectrum of observer qubit shown by stick diagram. Each transition of the spectrum correspond to the 
state of other qubits, which are given above each line.
(c) Deviation populations after creation of $\vert 00\rangle$ subsystem pseudopure state by POPS. 
Populations of only $\vert 00\rangle$  eigenstate is non-zero  in the two distinct domains of energy levels,
where observer qubit is respectively in state $\vert 0\rangle$ and $\vert 1\rangle$.
(d) Spectrum of observer qubit after creation of POPS. The dots denote null intensity.
(e) Deviation populations after a typical computation whose output is $\vert 11\rangle$. 
(f) Spectrum of observer qubit after such a computation. 

 Figure 3.  The quantum circuits of various algorithms.
(a) Quantum circuit for implementation of Grover's search algorithm in a 2-qubit system.
(b) Quantum circuit for implementation of approximate quantum counting in a 2-qubit system.
(c) Quantum circuit for implementation of Berstein-Vazirani problem.
(d) Quantum circuit for implementation of Hogg's algorithm in a 3-qubit system.

 Figure 4. (a) Chemical structure and equilibrium spectrum of 4-fluro 7-nitro benzofuran. The J-coupling values are 
$J_{01}$= -3.84 Hz, $J_{02}$=8.01 Hz and $J_{12}$= 8.07 Hz.
The peak denoted by asterisk (*) belongs to solvent. (b) Spectra after creation of POPS. A Gaussian shaped 
selective pulse of 500ms duration  was applied on the $\vert 000 \rangle \leftrightarrow \vert 100 \rangle$ transition and the 
resultant spectra is subtracted from the equlibrium spectra of figure (a) to yield (b). [See figure 2(c)] 

 Figure 5. (a) ``Spectral implementation" of Grover's search algorithm by 1D experiment. 
After computation, the observer qubit is detected by a non-selective pulse of 14 $\mu s$. 4$\times$1024 data points were collected 
and zero-filled to 8$\times$1024 before Fourier transform. The
observer qubit's spectra shows only the transition corresponding to the searched state ($\vert x\rangle$) with non-zero intensity.
(b) ``spectral implementation" of Grover's search algorithm by 2D experiment. A 2D data set of 256$\times$16 
($t_2\times t_1$)was collected and zero-filled to 1024$\times$256. It may be noticed that the total size of the raw 
2D dataset is of the same size as that of the 1D experiment. The doubly Fourier transformed spectra gives the input state 
along $\omega_1$ and output state along $\omega_2$.

 Figure 6. (a) ``Spectral implementation" of approximate quantum counting by 1D experiment.
 4$\times$1024 data points were collected and zero-filled to 8$\times$1024 before Fourier transform. The
observer qubit's spectra show the transitions corresponding to the ouput state. Hence the various cases of 
k=0 $(f_{00})$, k=1 $(f_{01}$ and  $f_{01})$, and k=2 $(f_{11})$ can be easily identified from the spectra.  
(b) ``spectral implementation" of approximate quantum counting by 2D experiment. A 2D data set of 256$\times$16
($t_2\times t_1$) was collected and zero-filled to 1024$\times$256. The Fourier transformed spectra gives the 
output state as well as the input state. 

 Figure 7. (a) ``Spectral implementation" of Berstein-Vazirani problem in a 2-qubit system.
 The observer qubit's spectra shows the transitions corresponding to the bit string. The strings 
$a=00$, $a=01$, $a=10$ and $a=11$ can be identified directly from the spectra.
(b) 2D ``spectral implementation" of Berstein-Vazirani problem. A 2D data set of 256$\times$16
($t_2\times t_1$)was collected and zero-filled to 1024$\times$256 before Fourier transform.
 The Fourier transformed spectra gives the bit string against the input state in each case.
 
 Figure 8. (a) Chemical structure and equilibrium spectrum of 4-fluro 7-nitro benzofuran. The J-coupling values are
$J_{01}$= 5.23 Hz, $J_{02}$= 8.85 Hz, $J_{03}$= 19.1 Hz, $J_{12}$= 9.76 Hz, $J_{13}$= -2.4 Hz 
and $J_{23}$= 6.81 Hz.
 (b) Spectra after creation of POPS. A Gaussian shaped
selective pulse of 500ms duration was applied on the $\vert 0000 \rangle \leftrightarrow \vert 1000 \rangle$ transition and the
resultant spectra is subtracted from the equlibrium spectra of figure (a) to yield (b).

Figure 9. (a) ``Spectral implementation" of Berstein-Vazirani problem in a 3-qubit system.
After computation, the observer qubit is detected by a selective pulse of 12.5us duration.
 The observer qubit's spectra show the transitions corresponding to the bit string. The 
eight possible strings of $a=000,a=001...a=111$ can be identified directly from the spectra.
(b) 2D ``spectral implementation" of Berstein-Vazirani problem in the three-qubit case. A 2D data set of 256$\times$24
($t_2\times t_1$) was collected and zero-filled to 1024$\times$256 before Fourier transform.
 The Fourier transformed spectra give the various bit strings along with the input state in each case.

Figure 10. One-dimensional ``spectral implementation" of Hogg's algorithm in a 3-qubit system.
After computation, the observer qubit is detected by a selective pulse of 12.5 $\mu s$ duration.
The observer qubits spectra clearly show the output states corresponding to various logical formulae of table 3.
(a) contains the spectra corresponding to m=1 and (b) to m=3.  

Figure 11. Two-dimensional ``spectral implementation" of Hogg's algorithm in a 3-qubit system.
The two-dimensional spectra provides the output states corresponding to various logical formulae of table 3.
(a) contains the spectra corresponding to m=1 and (b) to m=3.

\begin{figure}
\epsfig{file=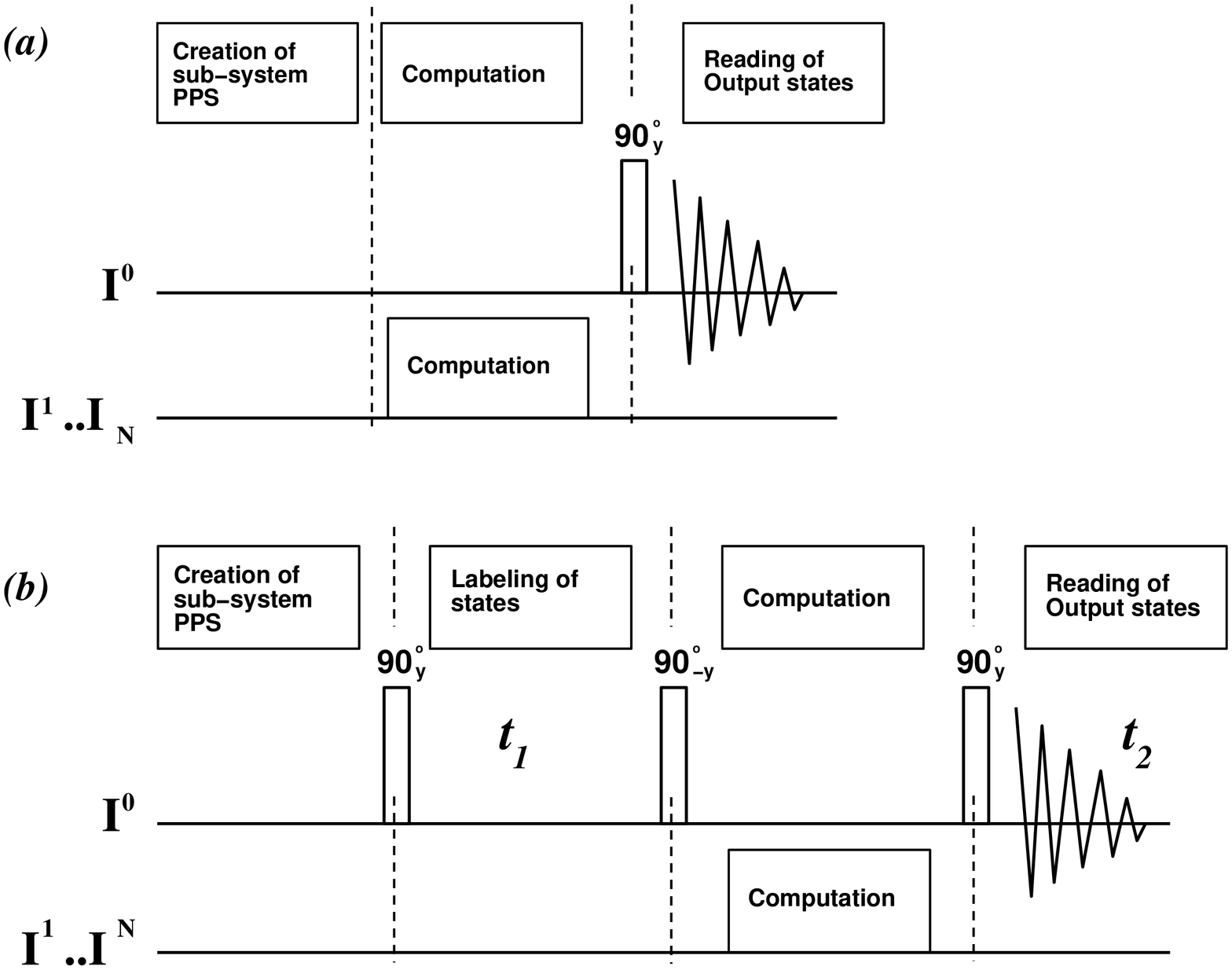,height=16cm,angle=270}
\end{figure}
\hspace{5cm}
{\huge Figure 1}
\pagebreak

\begin{figure}
\epsfig{file=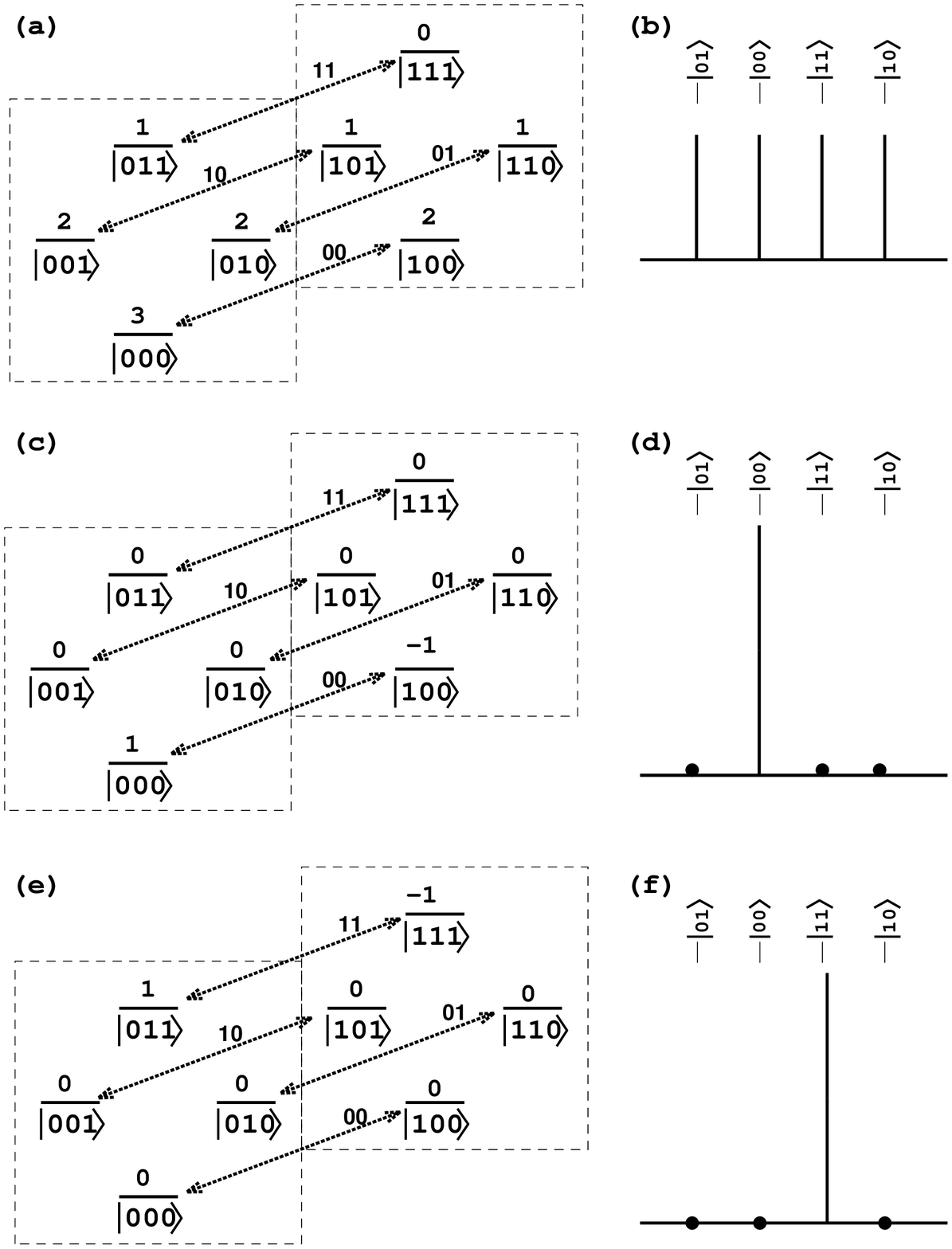,height=16cm}
\end{figure}
\hspace{5cm}
{\huge Figure 2}
\pagebreak

\begin{figure}
\epsfig{file=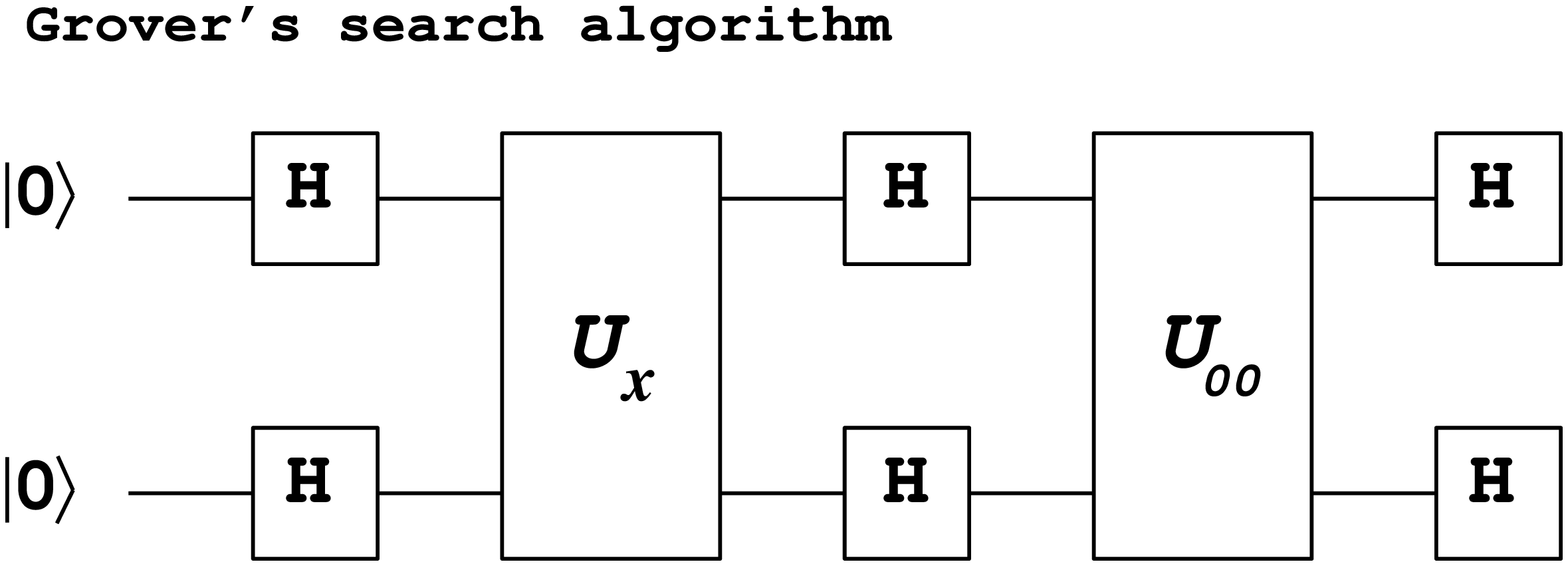,height=10cm,angle=270}
\end{figure}
\noindent {\vrule width 0.65\linewidth height 0.2pt depth 0.2pt}
\begin{figure}
\epsfig{file=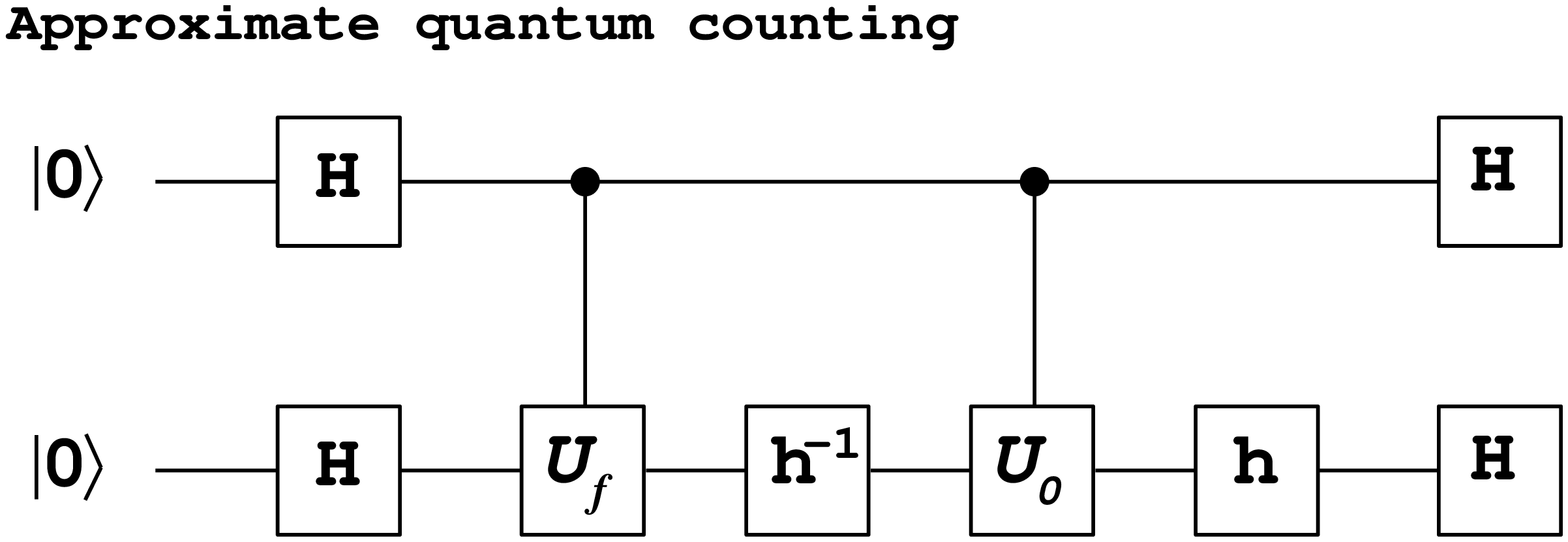,height=10cm,angle=270}
\end{figure}
\noindent {\vrule width 0.65\linewidth height 0.2pt depth 0.2pt}
\begin{figure}
\epsfig{file=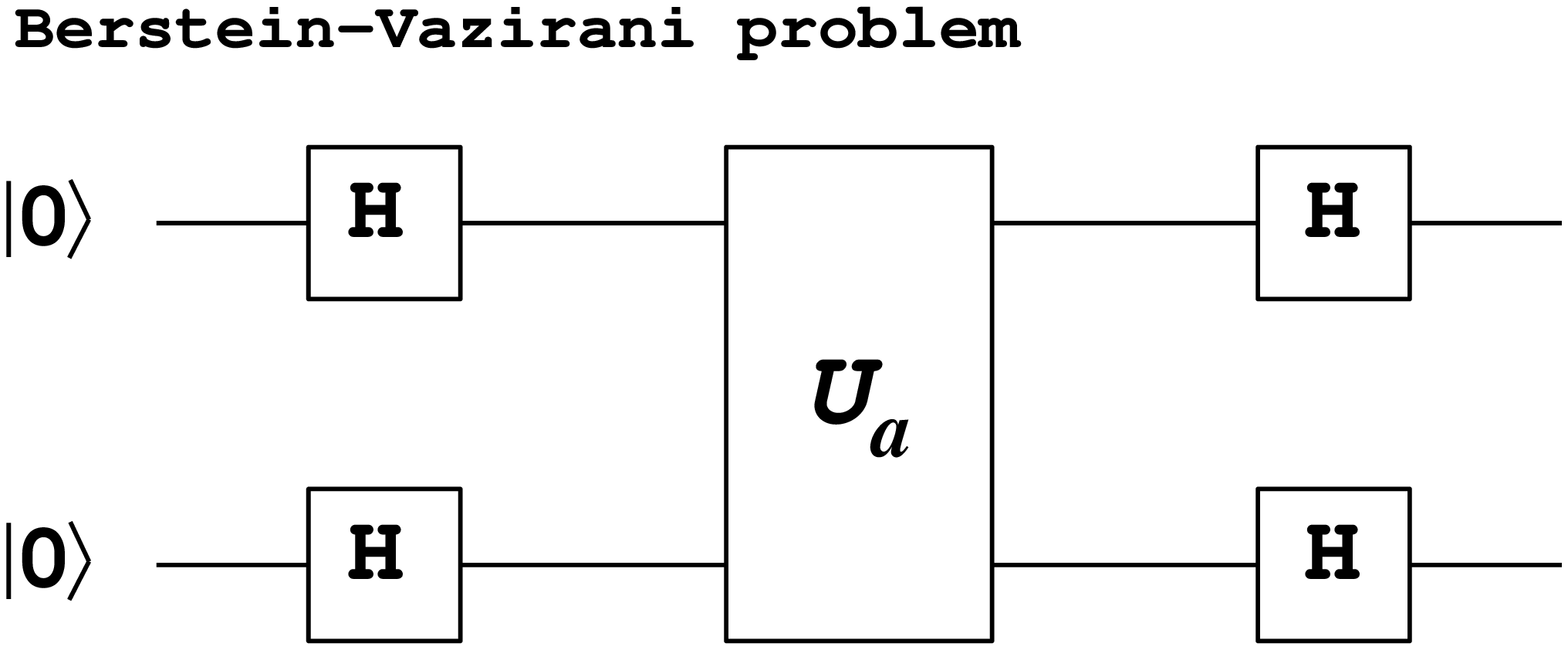,height=10cm,angle=270}
\end{figure}
\noindent {\vrule width 0.65\linewidth height 0.2pt depth 0.2pt}
\begin{figure}
\epsfig{file=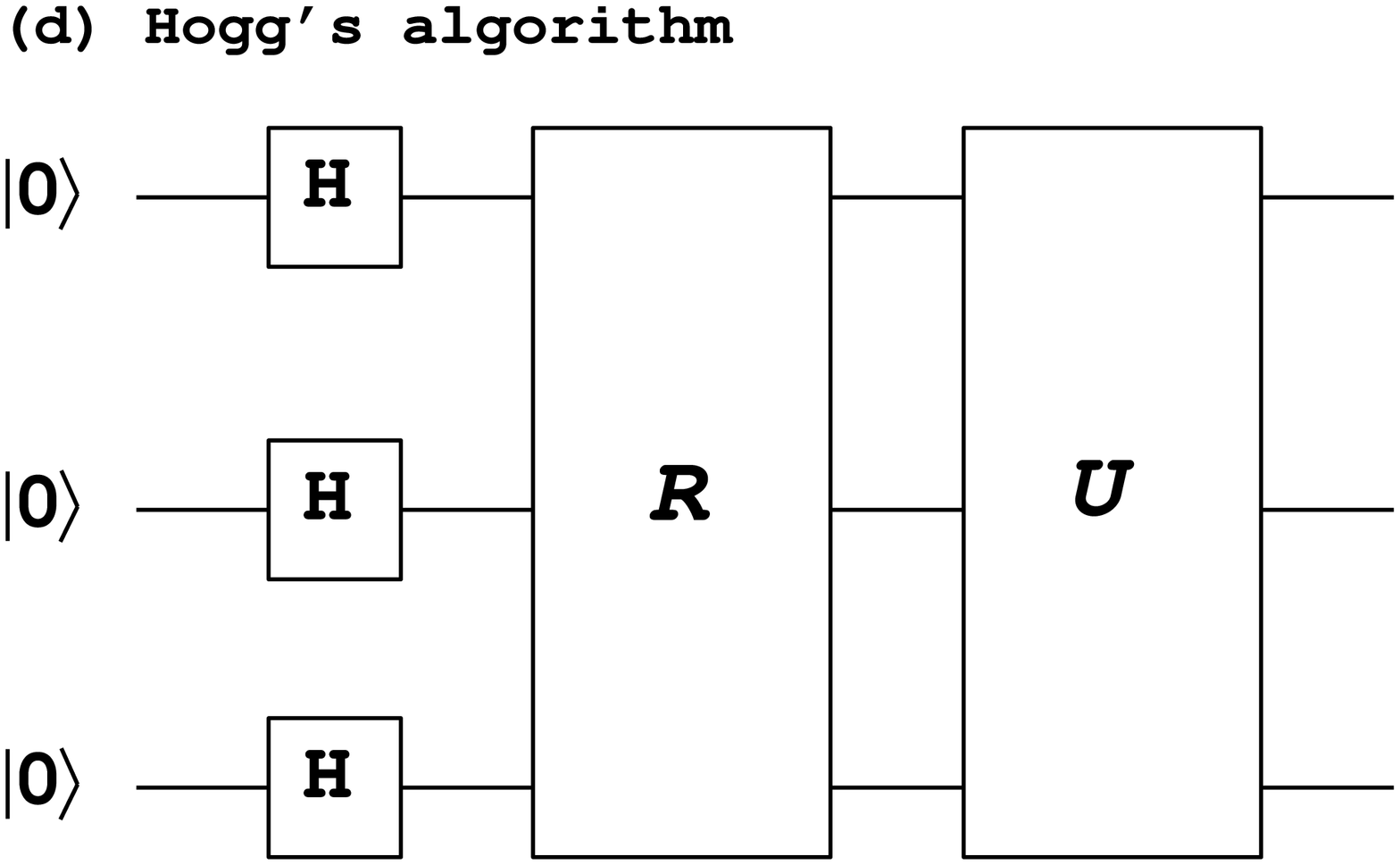,height=9.5cm,angle=270}
\end{figure}
\hspace{5cm}
{\huge Figure 3}

\begin{figure}
\epsfig{file=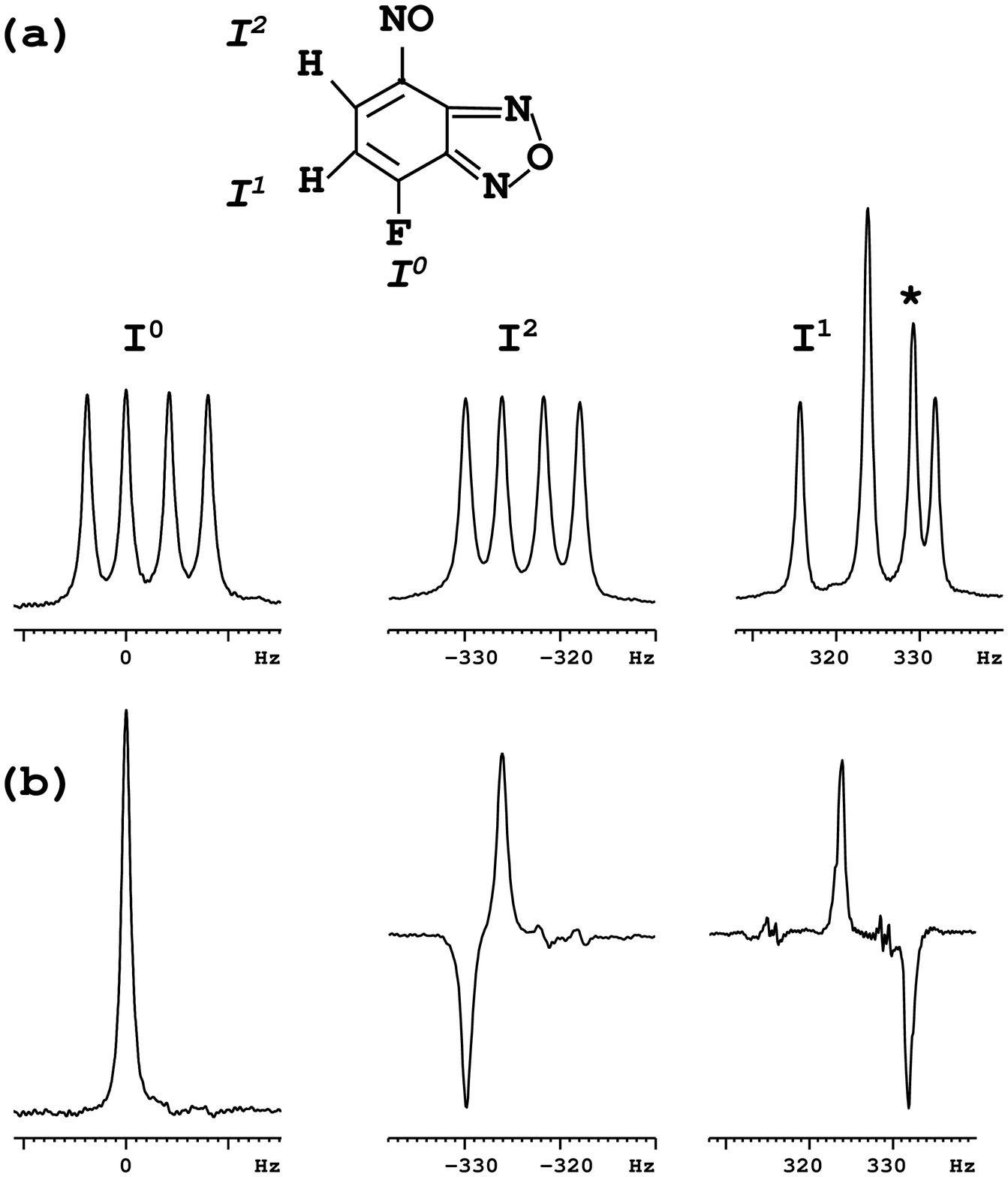,height=14cm}
\end{figure}
\hspace{5cm}
{\huge Figure 4}
\pagebreak

\pagebreak
{\Large (a)}
\vspace{-2.5cm}
\begin{figure}
\hspace{1cm}
\epsfig{file=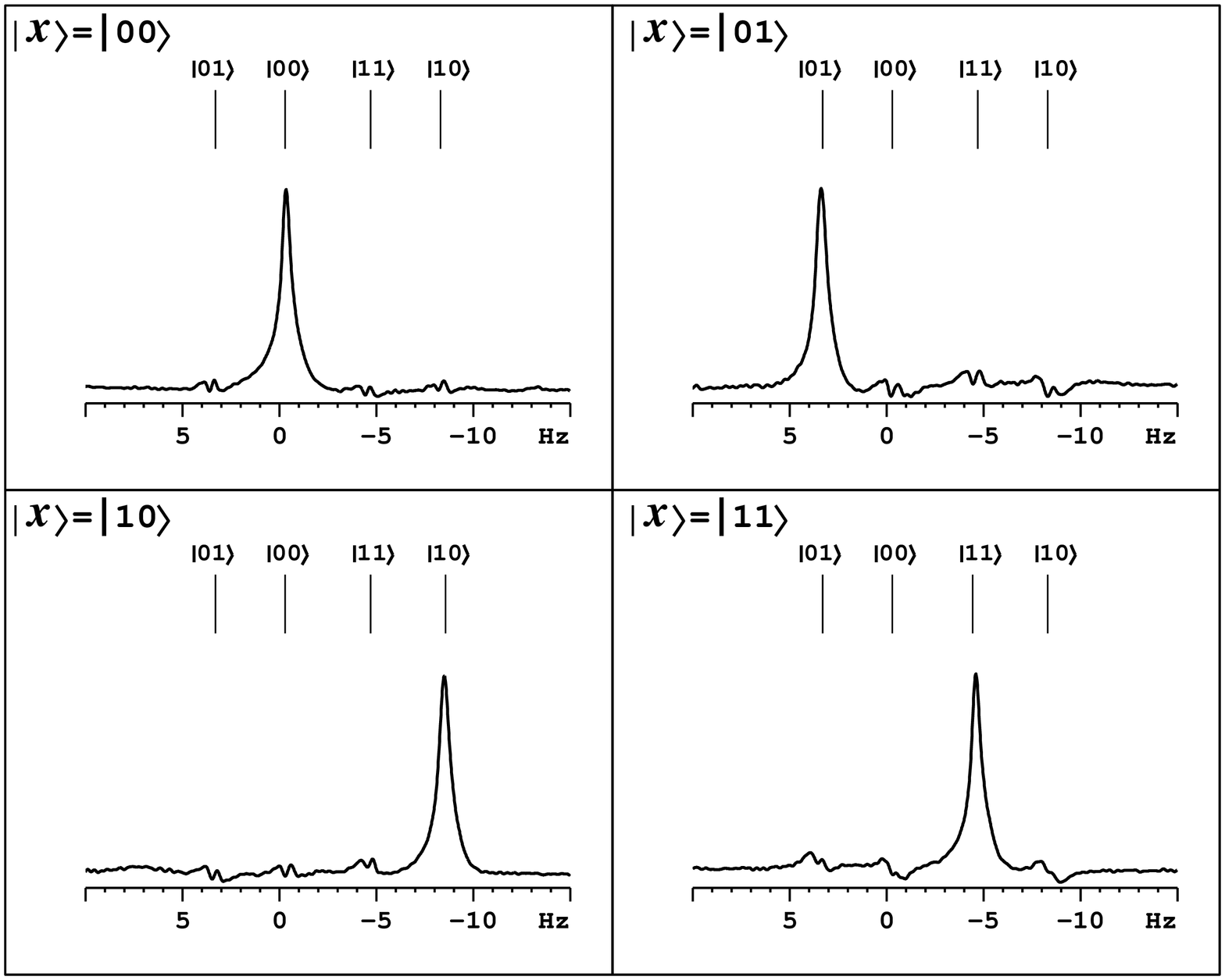,height=14cm,angle=270}
\end{figure}
\vspace{1cm}
{\Large (b)}
\vspace{-2cm}
\begin{figure}
\hspace{1cm}
\epsfig{file=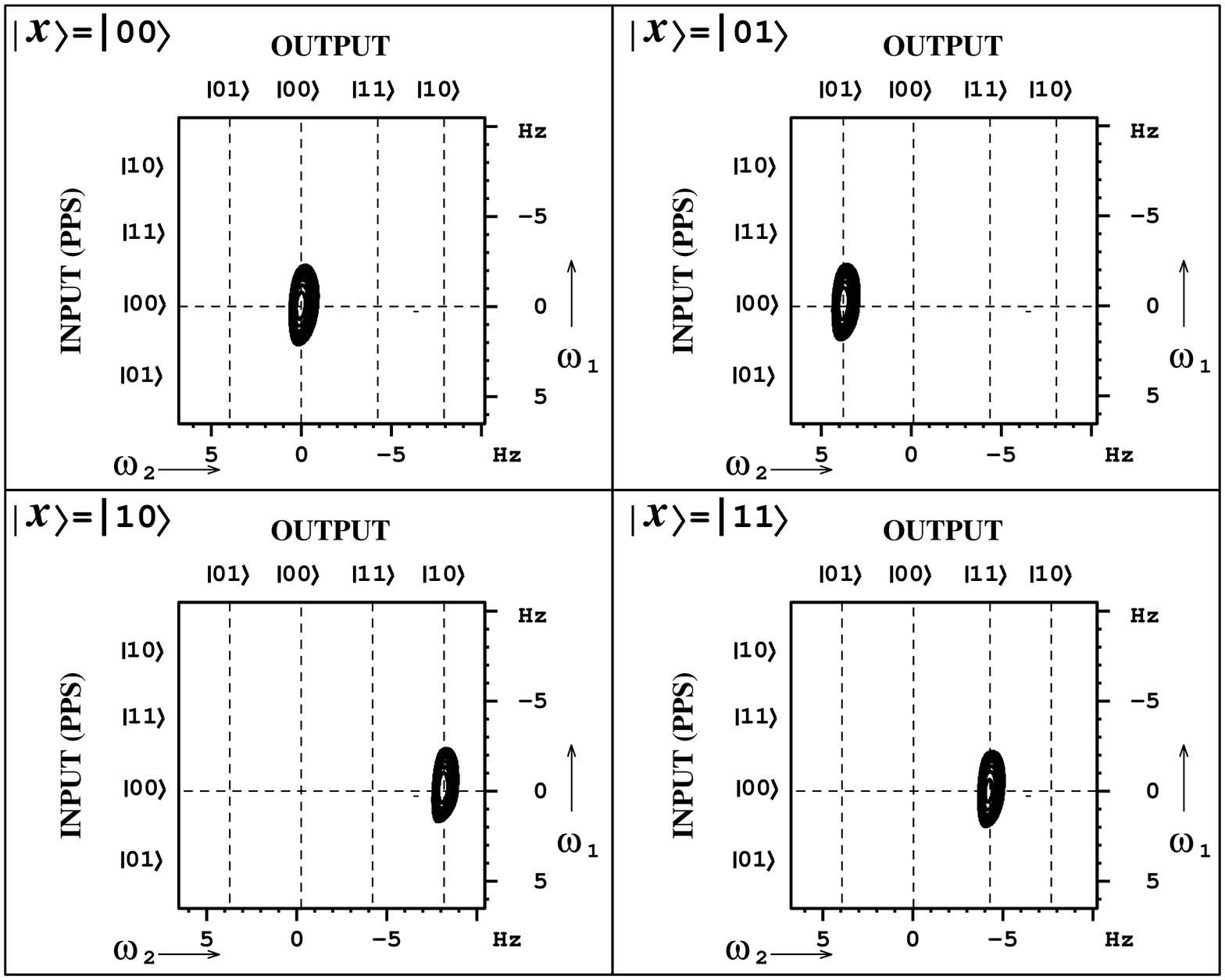,height=14cm,angle=270}
\end{figure}
\hspace{5cm}
{\Large Figure 5}

\pagebreak
{\Large (a)}
\vspace{-2.5cm}
\begin{figure}
\hspace{1cm}
\epsfig{file=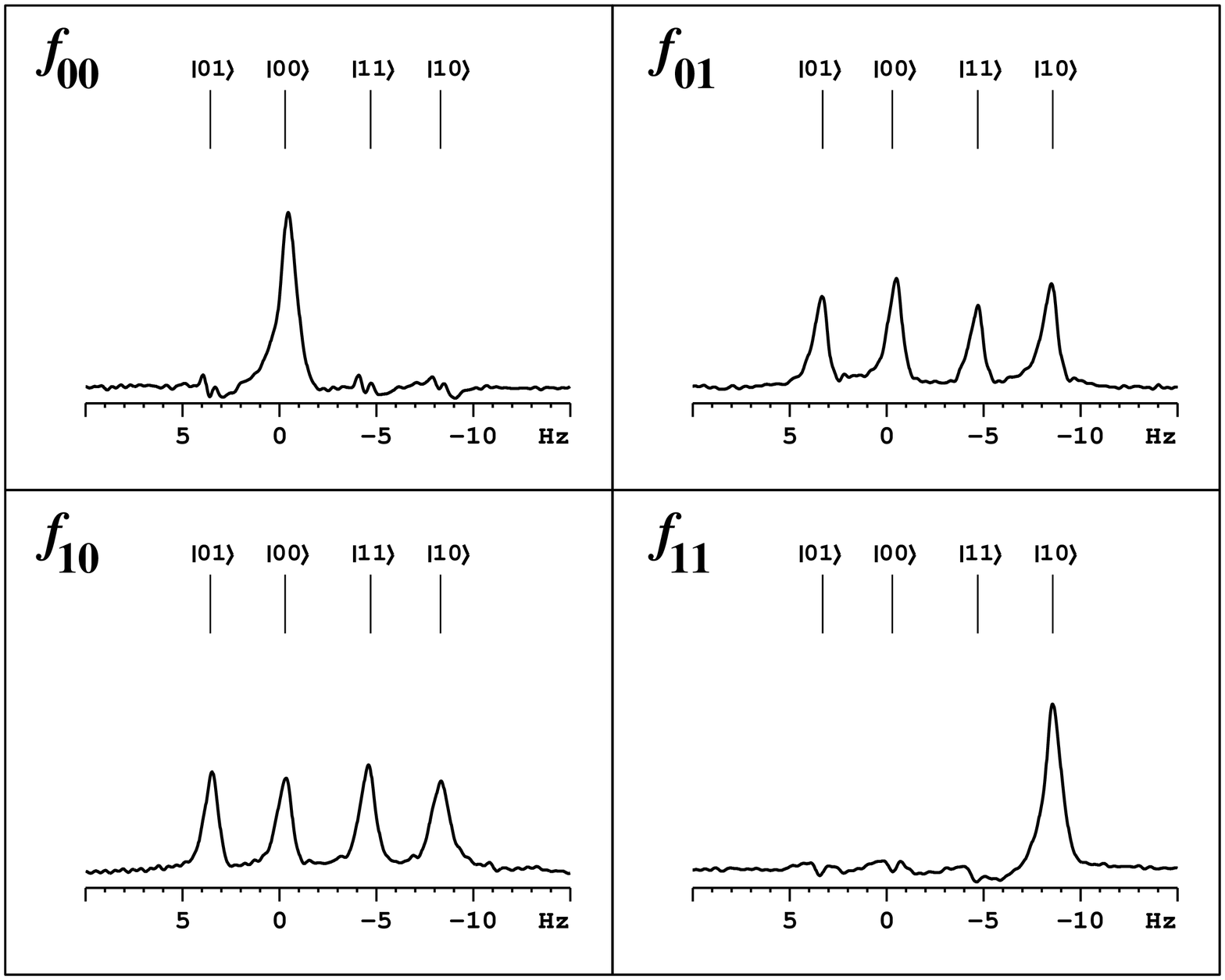,height=14cm,angle=270}
\end{figure}
\vspace{1cm}
{\Large (b)}
\vspace{-2cm}
\begin{figure}
\hspace{1cm}
\epsfig{file=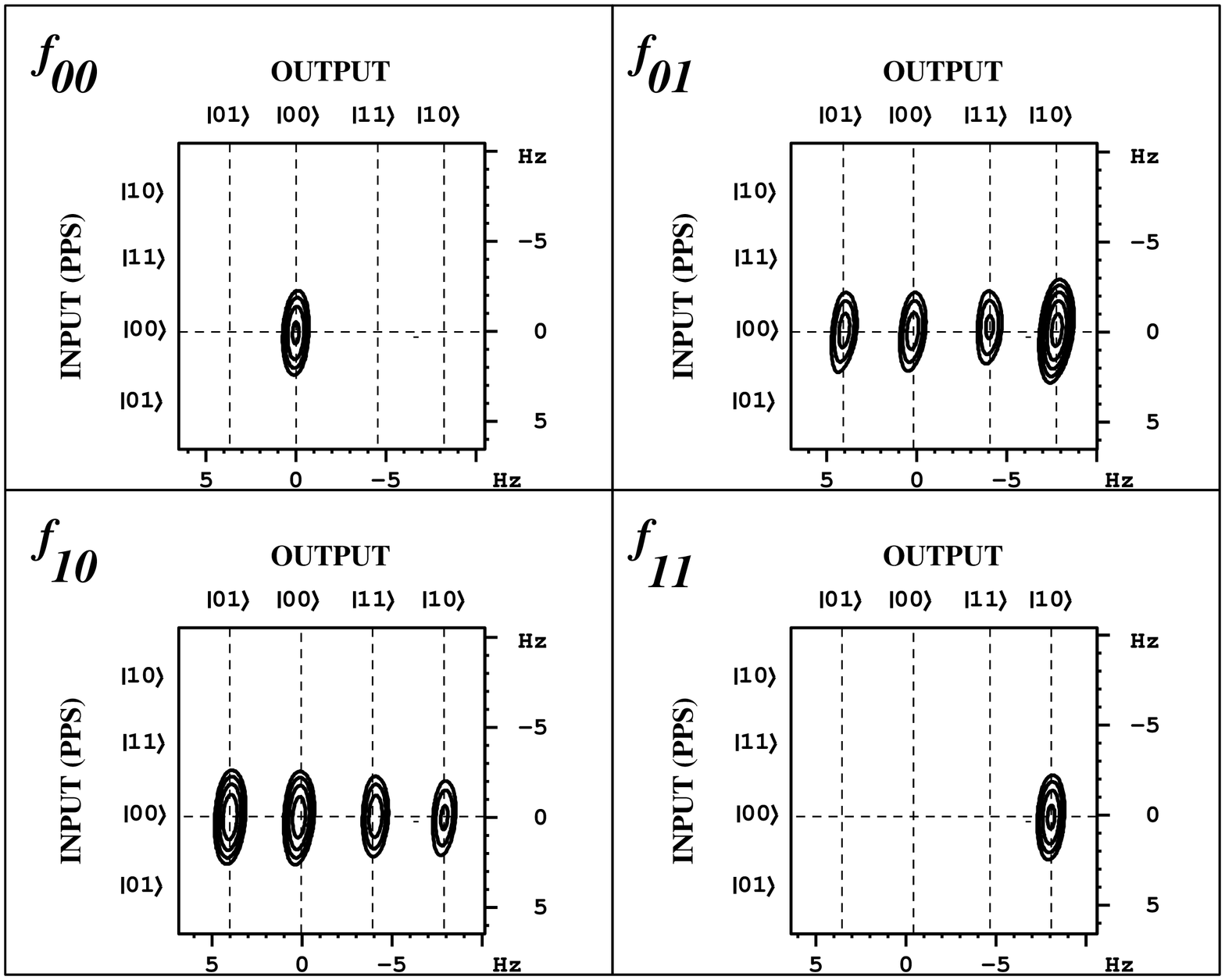,height=14cm,angle=270}
\end{figure}
\hspace{5cm}
{\Large Figure 6}
\pagebreak

{\Large (a)}
\vspace{-2.5cm}
\begin{figure}
\hspace{1cm}
\epsfig{file=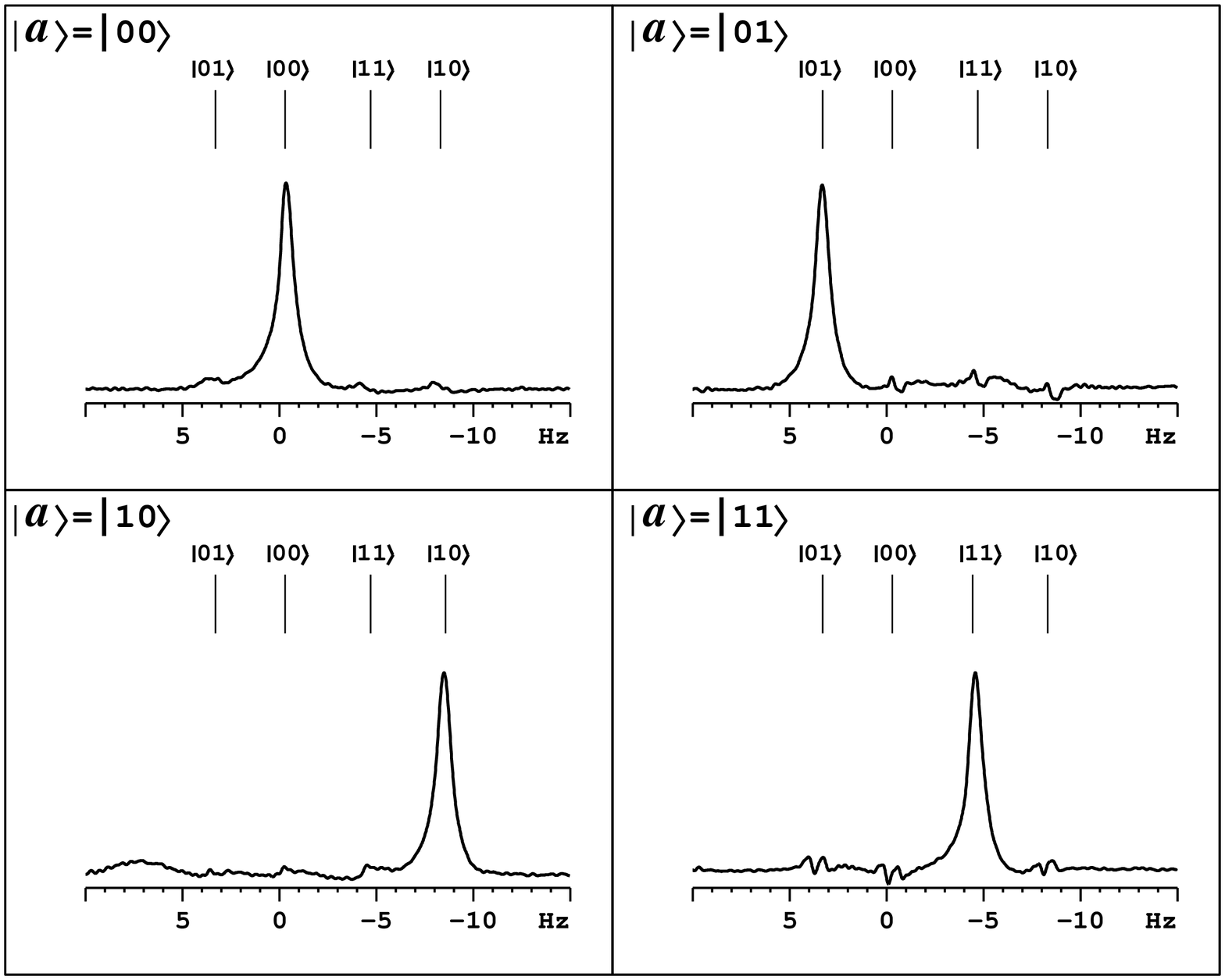,height=14cm,angle=270}
\end{figure}
\vspace{1cm}
{\Large (b)}
\vspace{-2cm}
\begin{figure}
\hspace{1cm}
\epsfig{file=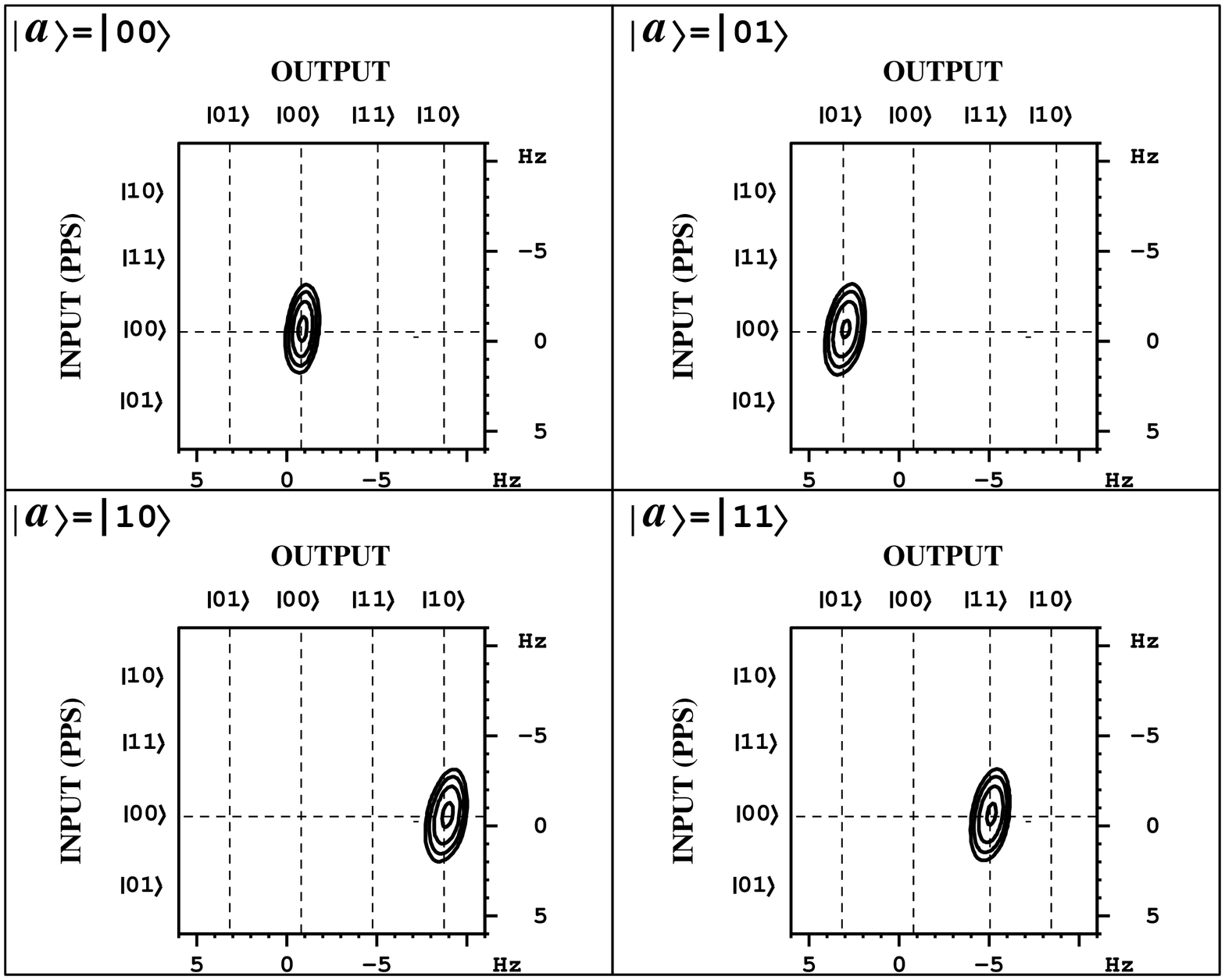,height=14cm,angle=270}
\end{figure}
\hspace{5cm}
{\Large Figure 7}

\pagebreak
\begin{figure}
\epsfig{file=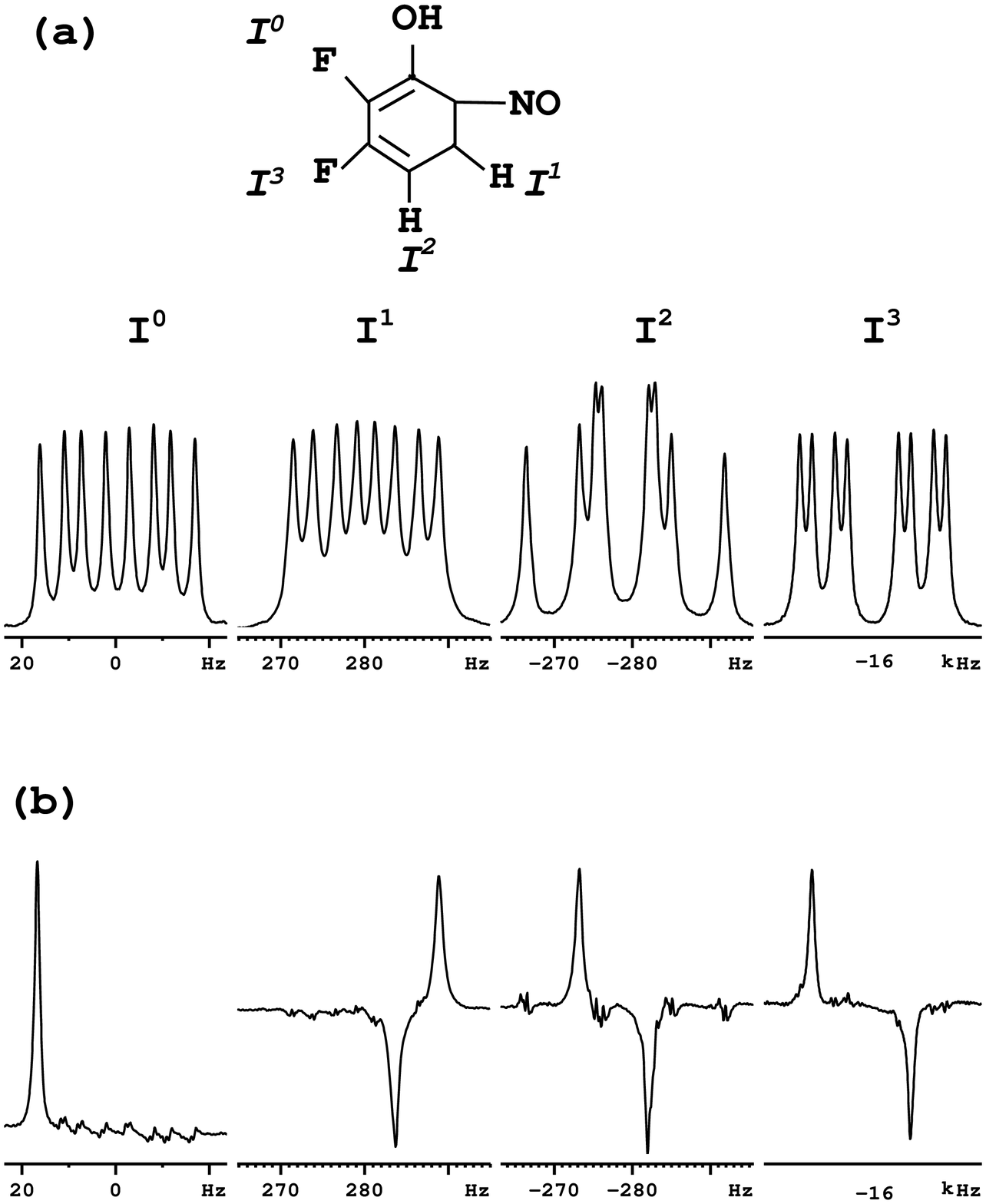,height=14cm}
\end{figure}
\hspace{5cm}
{\huge Figure 8}

\pagebreak
\begin{figure}
\epsfig{file=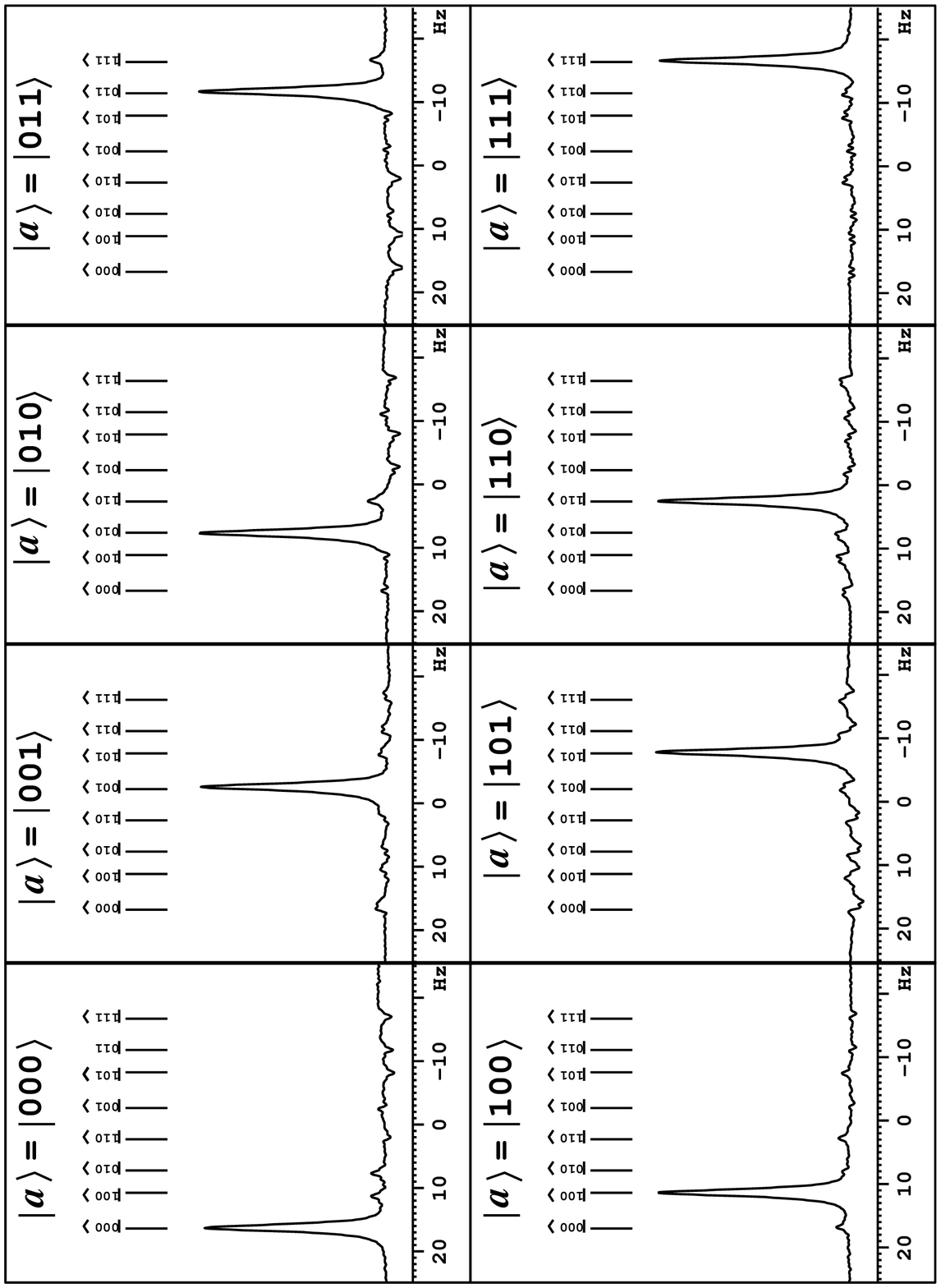,height=20cm}
\end{figure}
\hspace{7cm}
{ Figure 9(a)}
\begin{figure}
\epsfig{file=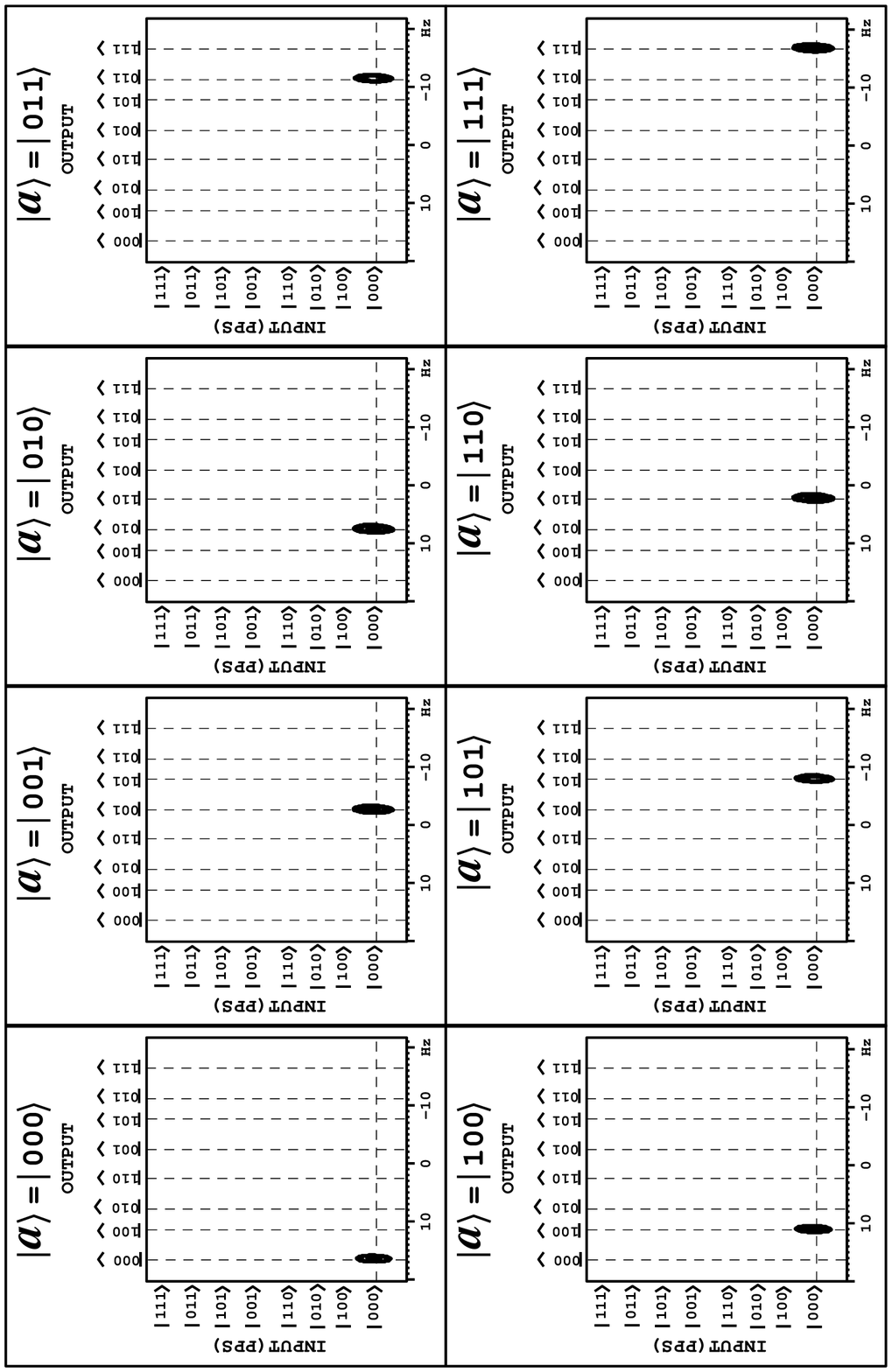,height=20cm}
\end{figure}
\hspace{7cm}
{Figure 9(b)}

\pagebreak
\begin{figure}
\epsfig{file=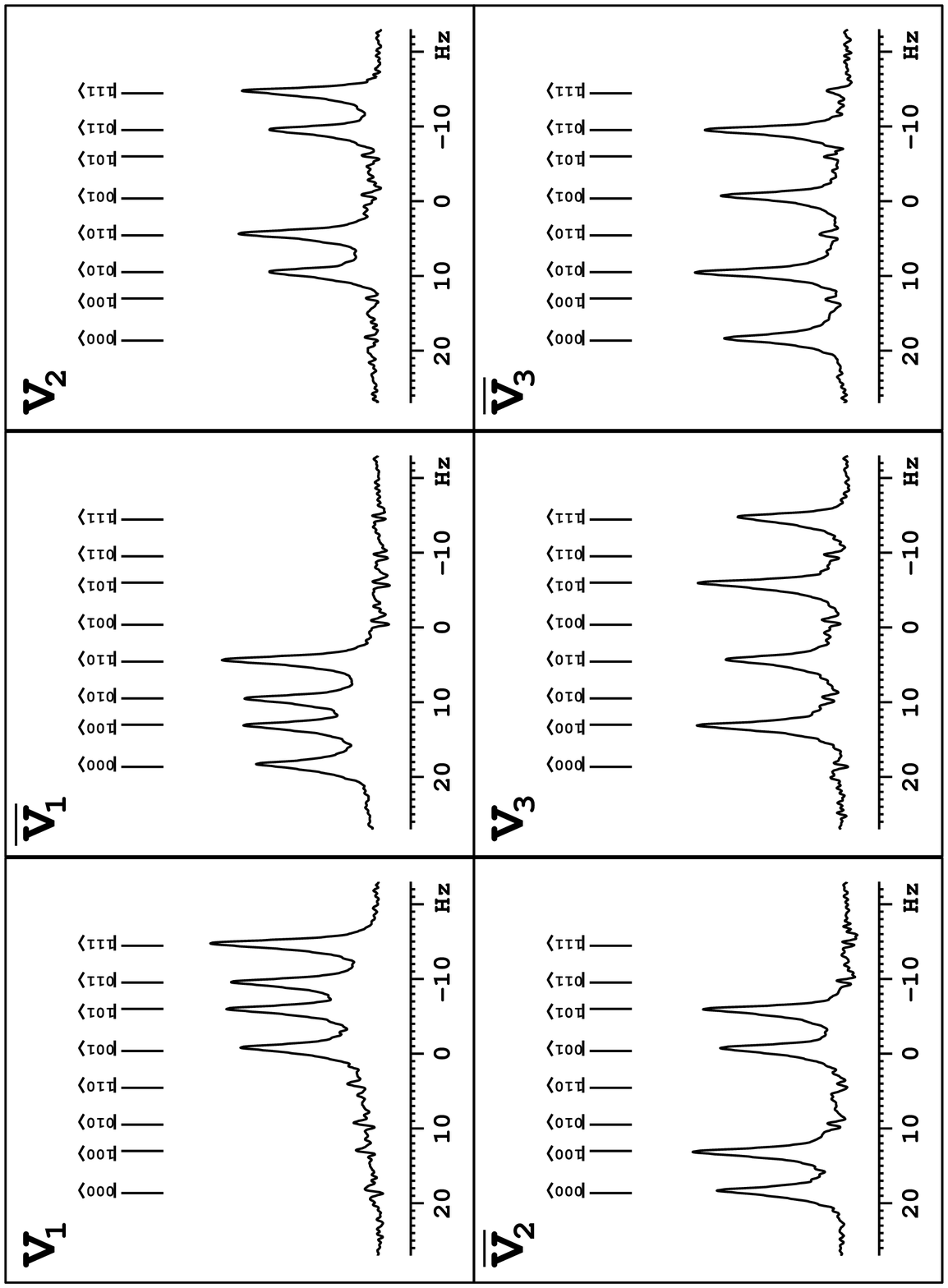,height=20cm}
\end{figure}
\hspace{7cm}
{Figure 10(a)}
\pagebreak
\begin{figure}
\epsfig{file=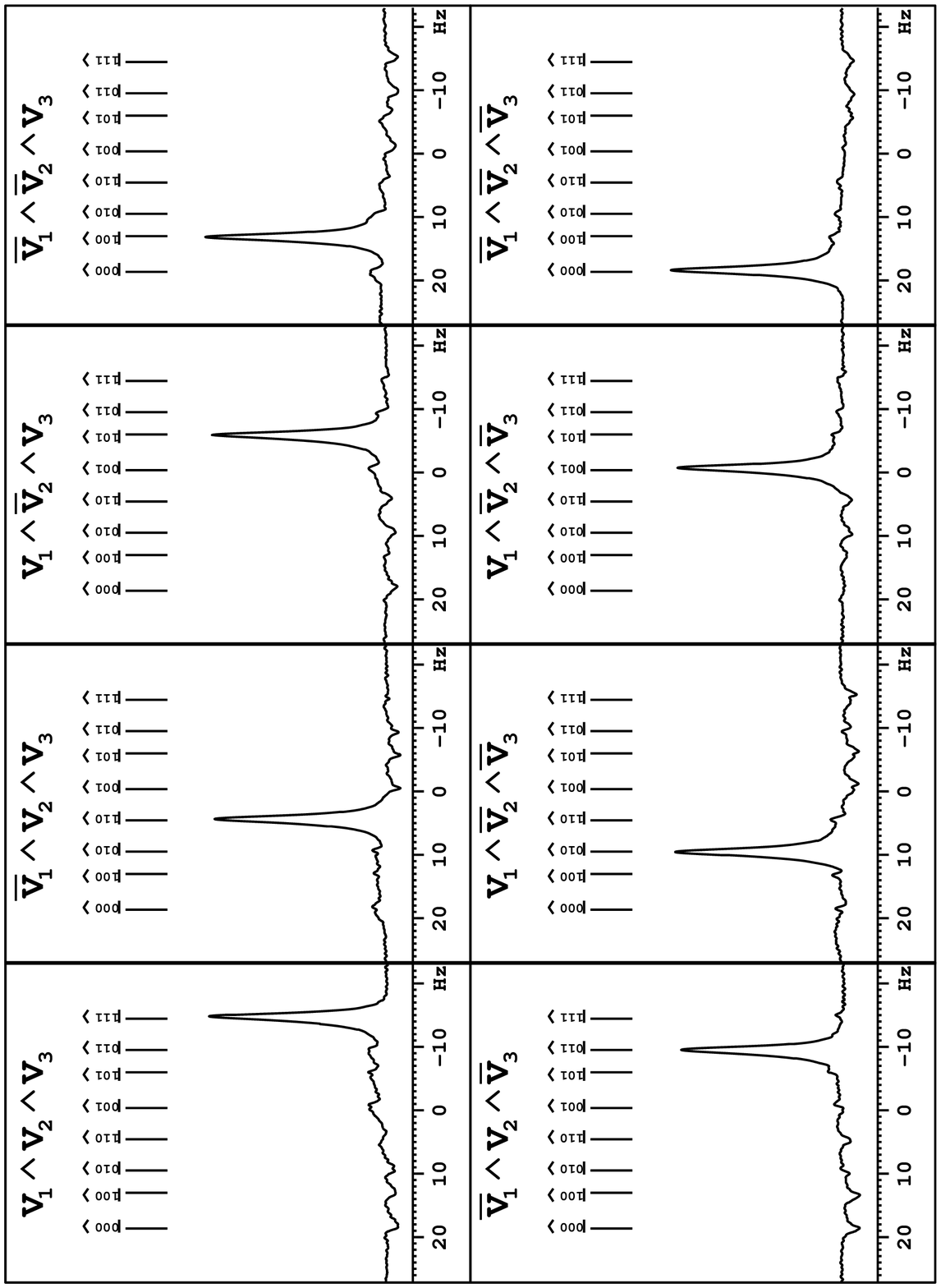,height=20cm}
\end{figure}
\hspace{7cm}
{Figure 10(b)}
\pagebreak
\begin{figure}
\epsfig{file=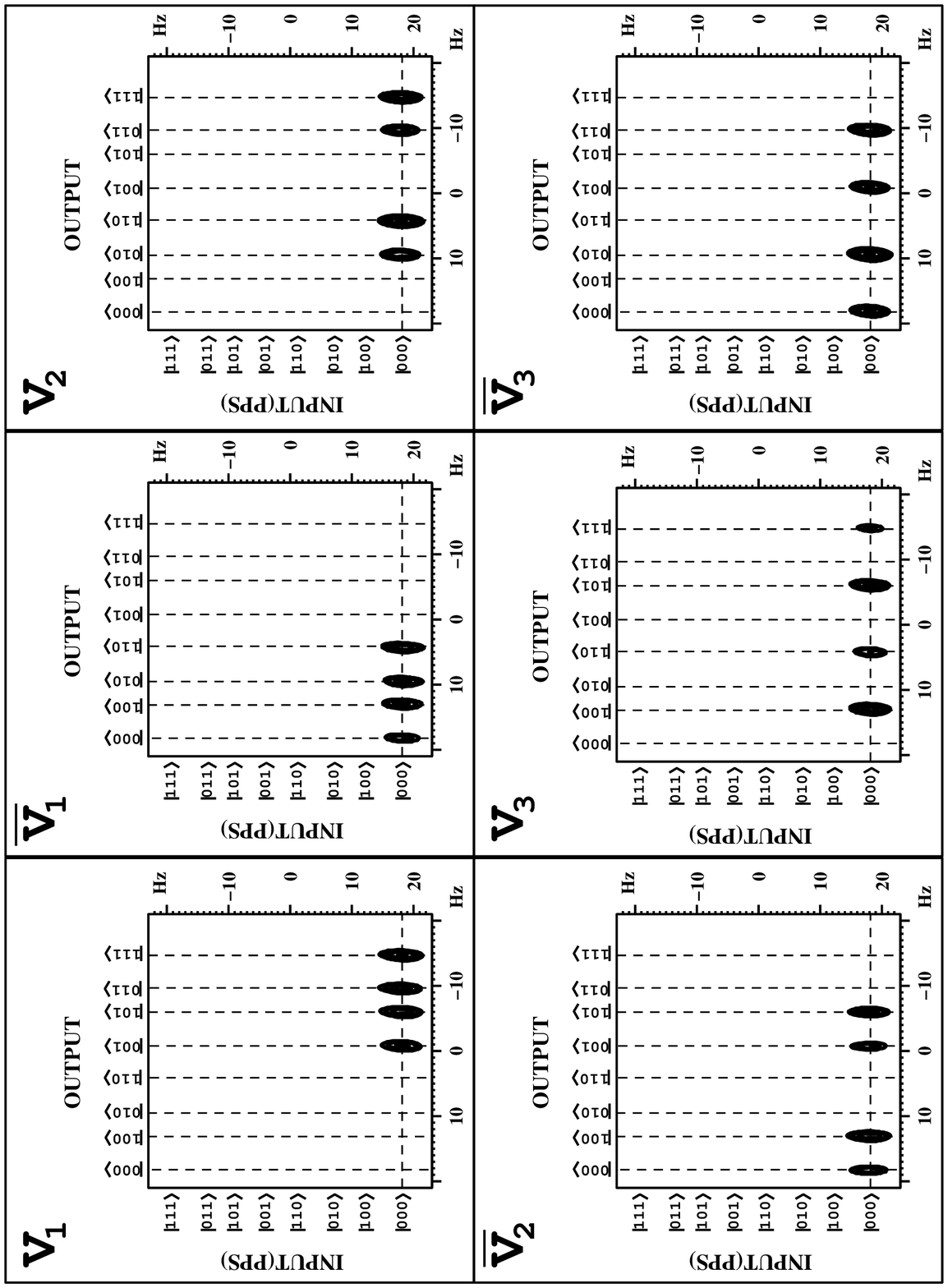,height=20cm}
\end{figure}
\hspace{7cm}
{ Figure 11(a)}
\pagebreak
\begin{figure}
\epsfig{file=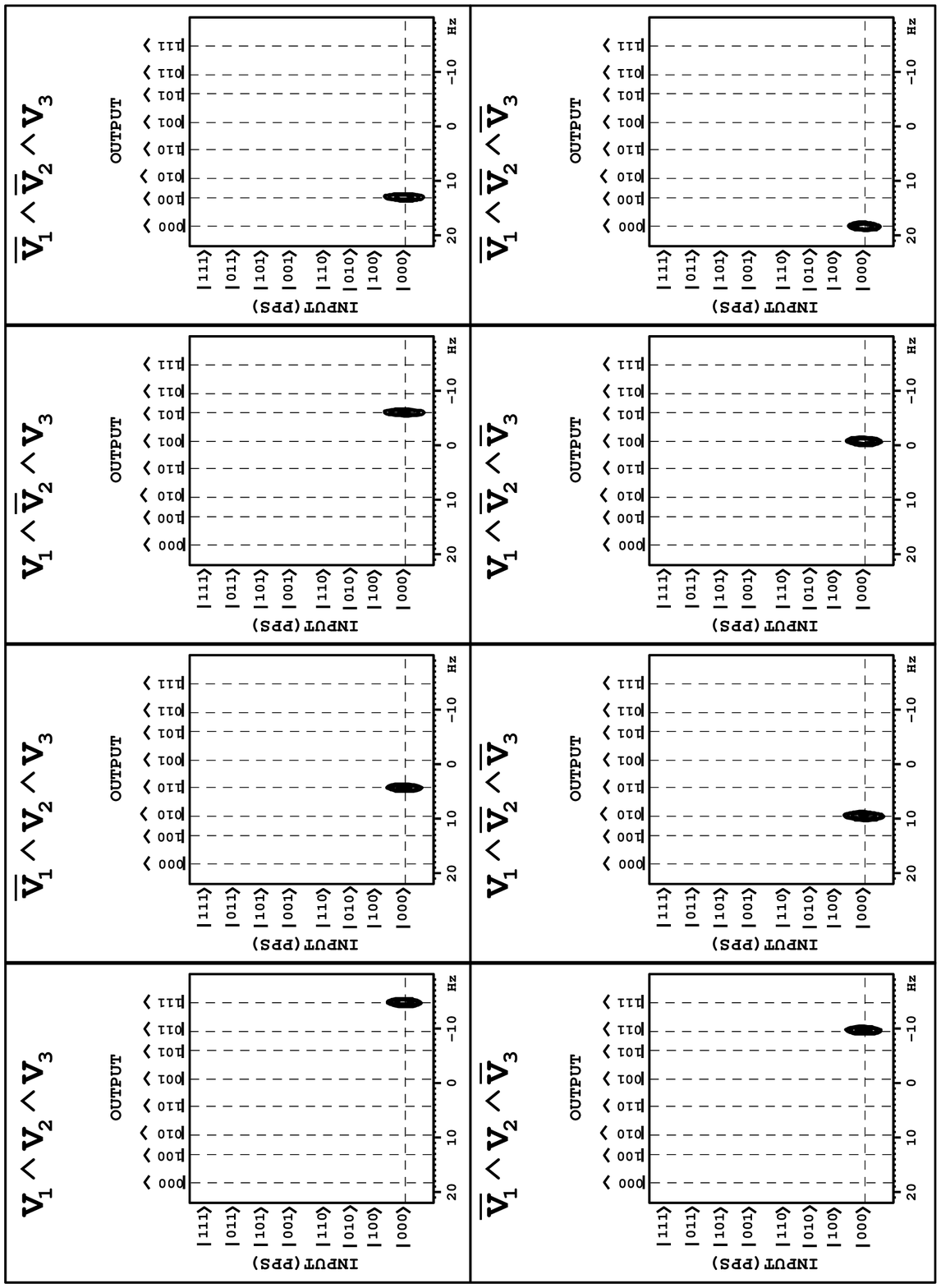,height=20cm}
\end{figure}
\hspace{7cm}
{ Figure 11(b)}


\begin{thebibliography}{99}
\bibitem{rf} R.P. Feynman, Int J. Theor. Phys. {\bf 21}, (1982) 467.
\bibitem{preskill} J. Preskill, {\it Lecture notes for Physics 229: Quantum information and Computation}, http://theory.caltech.edu/people/preskill/.
\bibitem{ss} S. Lloyd, Science {\bf 273}, 1073 (1996).
\bibitem{deu}  D. Deutsch and R. Jozsa,  Proc. R. Soc. Lond. A {\bf 400}, 97 (1985).
\bibitem{grover} L.K. Grover, Phys. Rev. Lett. {\bf 79}, (1997) 325.
\bibitem{shor} P.W.Shor, SIAM Rev. {\bf 41}, 303-332 (1999).
\bibitem{hogg} T. Hogg. Phys. Rev. Lett. {\bf 80}, 2473 (1998).
\bibitem{vazi} E. Berstein and U. Vazirani, SIAM J. Computin. {\bf 26}, 1411 (1997).  
\bibitem{count1} M. Boyer, G. Brassard, P. Hoyer, and A. Tapp. Fortschr. Phys. {\bf 46}, 493 (1998).
\bibitem{count2} G. Brassard, P Hoyer and A. Tapp, in Automata, Languages, and Programming: Proceedings of the 25$^{th}$
International Colloquium, ICALP'98, Aalborg, Denmark, 1998 (Springer, Berlin, 197); quant-ph/9805082.
\bibitem{bou} D. Bouwmeester, A. Ekert and A. Zeilinger, (Ed) {\it The Physics of Quantum Information}, Springer, 2000.
\bibitem{chuangbook} M.A. Nielsen and I.L. Chuang, {\it Quantum Computation and Quantum Information},
Cambridge University Press 2000.
\bibitem{cory97} D. G. Cory, A.F. Fahmy and T.F. Havel,  Proc.Natl.Acad.Sci. USA {\bf 94}, 1634 (1997).
\bibitem{chuang97} N. Gershenfeld and I.L. Chuang,   Science {\bf 275}, 350 (1997).
\bibitem{cory98} D. G. Cory, M. D. Price and T.F. Havel, Physica D, {\bf 120}, 82 (1998).
\bibitem{djchu} I. L.Chuang, L. M. K. Vanderspyen, X. Zhou, D.W. Leung, and S. Llyod,
  Nature (London) {\bf 393}, 1443 (1998).
\bibitem{djjo} J. A. Jones and M. Mosca,  J. Chem. Phys. {\bf 109}, 1648 (1998).
\bibitem{grochu} I. L. Chuang, N. Gershenfeld, and M. Kubinec, Phys. Rev. Lett. {\bf 80}, 3408-3411 (1998).
\bibitem{grojo} J. A. Jones, M. Mosca and R. H. Hansen, Nature {\bf 393} 344 (1998).
\bibitem{ka1} Kavita Dorai, T. S. Mahesh, Arvind and Anil Kumar, Current Science, {\bf 79} (10) 1447 (2000).
\bibitem{ka} Kavita Dorai, Arvind, Anil Kumar, Phys Rev A. {\bf 61}, 042306 (2000).
\bibitem{jcp} N. Sinha, T.S. Mahesh, K.V. Ramanathan and Anil Kumar, J. Chem. Phys. {\bf 114}, 4415 (2001).
\bibitem{pram} Arvind, K. Dorai and Anil Kumar, Pramana {\bf 56} 7705 (2001).  
\bibitem{nat} L.M.K. Vanderspyen, Matthias Steffen, Gregory Breyta, C.S.Yannoni,
 M.H. Sherwood and I.L. Chuang,   Nature {\bf 414}, 883 (2001).
\bibitem{pram1} Anil Kumar, K. V. Ramanathan, T. S. Mahesh, N. Sinha and K.V.R.M. Murali, Pramana {\bf 59} 243 (2002).
\bibitem{ranapra2} Ranabir Das and Anil Kumar,  Phys. Rev. A. {\bf 68}, 032304 (2003). 
\bibitem{ijqi} Ranabir Das, A. Mitra, S. Vijaykumar and Anil Kumar Int. Jour. of Quant. Infor. 1(3) 387 (2003).
\bibitem{ranapra1} Ranabir Das, Sukhendu Chakraborty, K. Rukmani and Anil Kumar,  Phys. Rev. A. (in press).
\bibitem{jones} J. A. Jones, chapter 5, page 181 of reference \cite{bou}.
\bibitem{chutomo} I.L. Chuang, N. Greshenfeld, M.Kubinec, and D. Leung, Proc. R. Soc. Lond. A {\bf 454}, 447 (1998).
\bibitem{ranabirtomo} Ranabir Das, T.S. Mahesh and Anil Kumar, Phys. Rev. A. {\bf 67}, 062304 (2003).
\bibitem{newtomo} Garett M. Leskowitz and Leonard J. Mueller, Phys. Rev. A. {\bf 69}052302 (2004). 
\bibitem{er} Z.L. Madi, R. Bruschweiler and R.R. Ernst, J. Chem. Phys. {\bf 109}, 10603 (1998).
\bibitem{lo} T. S. Mahesh, Kavita Dorai, Arvind, Anil Kumar, J. Mag. Res. {\bf 148}, 95 (2001).
\bibitem{spec1}  X. Peng, X. Zhu, X. Fang, M. Feng, X. Yang, M. Liu, and K. Gao,  quant-ph/0202010.
\bibitem{spec2}  X. Peng, X. Zhu, X. Fang, M. Feng, M. Liu, and K. Gao, {\it J. Chem. Phys.} {\bf 120}, 3579 (2004).
\bibitem{ernstbook} R.R. Ersnt, G. Bodenhausen, and A. Wokaun, {\it Principles of Nuclear Magnetic Resonance
 in One and Two Dimensions}, Oxford university press 1987.
\bibitem{fung00} B. M. Fung, Phys. Rev. A {\bf 63}, 022304 (2001).
\bibitem{fungjcp} Vladimir L. Ermakov and B. M. Fung, J. Chem. Phys. {\bf 118}, 10376 (2003).
\bibitem{cat} E. Knill, R. Laflamme, R. Martinez, and C.-H. Tseng, Nature (London) 404, 368 (2000).
\bibitem{levitt} R. Freeman, T.A. Frenkiel and M.H. Levitt, {\it J. Mag. Res.}{\bf 44}, 409(1981).
\bibitem{jump} P. Plateau and M. Guéron, J. Am. Chem. Soc. {\bf 104}, 7310 (1982).
\bibitem{nmrcount} J. A. Jones amd M. Mosca, Phys. Rev. Lett. {\bf 83} 1050 (1999).
\bibitem{nmrcloning} H. K. Cummins C. Jones A. Furze, N. F. Soffe, M. Mosca, J. M. Peach and J. A. Jones, Phys. Rev. Lett
{\bf 88} 187901 (2002).
\bibitem{nmrhogg} X. Peng, X. Zhu, X. Fang, M. Feng, M. Liu, and K. Gao, Phys. Rev. A. {\bf 65} 042315 (2002).
\bibitem{nmrvazi} J. Du, M. Shi, X. Zhou, Y. Fan, B. Ye, and R. Han, Phys. Rev. A. {\bf 64} 042306 (2001). 
\end{thebibliography}
\end{document}